\documentclass[submission,copyright,creativecommons]{eptcs}
\usepackage{amsmath}
\usepackage{graphicx}
\graphicspath{{./diagrams/}}
\usepackage{subcaption}
\usepackage{color}
\usepackage{soul}

\title{On the Learnability of Programming Language Semantics}
\author{
	Dan R. Ghica
	\institute{University of Birmingham\\  Birmingham, UK}
	\email{D.R.Ghica@cs.bham.ac.uk}
	\and 
	Khulood Alyahya
	\institute{University of Exeter\\  Exeter, UK\\
	King Saud University\\ Riyadh, KSA}
    \email{K.Alyahya@exeter.ac.uk}
    \email{KAlyahya1@ksu.edu.sa}
}

\def\titlerunning{On the Learnability of Programming Language Semantics}
\def\authorrunning{D. R. Ghica and K. Alyahya}
\begin{document}
\maketitle
\begin{abstract}
	Game semantics is a powerful method of semantic analysis for programming languages. It gives mathematically accurate models (``fully abstract") for a wide variety of programming languages. Game semantic models are combinatorial characterisations of all possible interactions between a term and its syntactic context. Because such interactions can be concretely represented as sets of sequences, it is possible to ask whether they can be learned from examples. Concretely, we are using long short-term memory neural nets (LSTM), a technique which proved effective in learning natural languages for automatic translation and text synthesis, to learn game-semantic models of sequential and concurrent versions of Idealised Algol (IA), which are algorithmically complex yet can be concisely described. We will measure how accurate the learned models are as a function of the degree of the term and the number of free variables involved. Finally, we will show how to use the learned model to perform latent semantic analysis between concurrent and sequential Idealised Algol.
\end{abstract}

\section{Programming languages and machine learning}

Programming language semantics, the way we ascribe meaning to programming languages, comes in different flavours. There is the operational approach, which consists of a collection of effective syntactic transformations that describes the execution of the program in a machine-independent way (see~\cite{DBLP:conf/ac/Pitts00} for a tutorial introduction). There is also the denotational approach, in which subprograms (terms) are interpreted, compositionally on syntax, as objects in a mathematical semantic domain (see~\cite{tennent1976denotational} for an introduction). The two approaches are complementary, and both have been studied extensively. Most commonly, especially for most ``real life'' programming languages, there is another, \textit{ad hoc}, approach of specifying a language, through a compiler, often informally described in a ``standard''.

Relating operational and denotational models is a mathematically difficult but worthwhile endeavour. Term equality is operationally defined in a way which is almost unworkable in practice: contextual equivalence. By contrast, term equality in the denotational model is just equality of the denoted mathematical objects. As a result denotational models are presumably handy in applications where equality of terms is important, such as compiler optimisations. When contextual equivalence coincides with semantic equality the model is said to be \textit{fully abstract}, a gold standard of precision for a denotational model. Constructing fully abstract denotational models even for relatively simple higher order (PCF~\cite{DBLP:journals/tcs/Plotkin77}) or procedural (Algol~\cite{o2013algol}) languages turned out to be a difficult problem, extensively studied in the 1990s. Many interesting semantic developments emerged out of this concerted effort, including \textit{game semantics}, a technique which finally gave the first such fully abstract models first for PCF~\cite{DBLP:journals/corr/AbramskyJM13} and Algol~\cite{DBLP:journals/entcs/AbramskyM96} then for many other programming languages~\cite{DBLP:conf/lics/Ghica09}.

Relating any mathematical (operational or denotational) model to the de facto ``model'' which is the compiler is a much different proposition. Whereas constructing a compiler from a mathematical specification is an arduous but achievable task, what we want to consider is the converse question.
Given a compiler, could we, at least in principle, construct a semantic model of the language? What is the right avenue of attack for this daunting problem?

A compiler is in some sense a formal specification. However, the compiler as a specification does not help us reason about basic properties of terms, such as contextual equivalence. How can we extract a more conventional kind of semantics? On the face of it, the question may seem preposterous at worst, unanswerable at best. 

Operational semantics (OS), the workhorse of much applied programming language theory, seems an unsuitable candidate for this job. Much like structural models of natural language, the rules of OS have a syntactic intricacy which cannot be hoped to be reconstructed from behavioural observations. Noting that recent progress has been achieved in learning the structural semantics of natural languages~\cite{DBLP:conf/iwcs/GrefenstetteDZS13}, operational semantics of programming languages cannot take advantage of these methods. For example the basic beta-reduction rule present in some form in all functional languages requires a complex form of substitution which assumes the concepts of binder, free variable and alpha equivalence. Denotational semantics on the other hand seems a more plausible candidate because of its independence of syntax. A final key observation is that some denotational models can be mathematically elementary. This is true of trace-like models in general~\cite{DBLP:journals/jcss/FrancezHLR79} and game semantic models~\cite{abramsky1997semantics} in particular. In fact one can think of game semantics as compositional trace models suitable for higher order programming languages. This seems to give us a foothold in attacking the problem. If a model can be specified simply as a set of traces subject to combinatorial constraints, perhaps such models can be machine-learned using techniques that proved successful in the learning of natural languages.

For tutorials and surveys of game semantics the reader is referred to the literature~\cite{abramsky1999game,DBLP:conf/lics/Ghica09}. The basic elements of a game semantics are \textit{moves}, with a structure called an \textit{arena}. Arenas are determined by the type signature of the term and consist of all the possible interactions (calls and returns) between a term and its context. Sequences of interactions are called \textit{plays} and they characterise particular executions-in-context. Finally, terms are modelled by sets of plays called \textit{strategies}, denoting all possible ways in which a term can interact with its context.
 
Certainly, not all interactions are possible, so plays are constrained by legality conditions. Conversely, strategies are subject to certain closure conditions, such as prefix-closure, stipulating that if certain plays are included so must be other ones. Because all features of a game semantic model are combinatorial properties of sequences (plays) or sets of sequences (strategies), using machine learning to identify them is no longer a preposterous proposition. The question certainly remains whether these properties can be learned and how accurately. 

In this paper we present two sets of computational experiments focussing on the learnability of known game semantic models of two similar programming languages. We first look at the intrinsic learnability of the language, automatically creating models from positive examples of legal plays, tested against sets of plays which are slightly modified so that to become illegal. The second experiment uses the learned models to perform latent semantic analysis~\cite{landauer2006latent} on the two languages, attempting to determine the provenance of a set of legal plays.

To learn the model we use neural networks, more precisely \textit{long short-term memory neural nets}~\cite{DBLP:journals/neco/HochreiterS97} (LSTM), which proved to be highly successful in automated translation~\cite{wu2016google} and text synthesis~\cite{fan2014tts}. The results are surprisingly good, with the trained net being able to reliably discriminate both between legal and illegal plays, and between legal plays from two slightly different programming languages. Moreover, the neural net had a standard architecture and, relative to LSTMs used in natural language processing, was quite small. Training converged rapidly, requiring relatively modest computational resources. 

These positive results should be received with cautious optimism. Methodologically, a strong case can be made that game semantics gives a possible angle of attack on the machine-learning problem for programming languages, compared to operational or other denotational programming language semantics (e.g. domain-theoretic~\cite{abramsky1994domain}). Moreover, it appears that the algorithmically complex combinatorial patterns which characterise the legality of game models are learnable enough to be able to reliably distinguish between legal plays and plays with small illegal irregularities and between plays belonging to slightly different languages. 

Of course, the resulting model is opaque and cannot serve as a basis for true understanding of a language, but it could be the starting point of a deeper automation of certain programming language processes which require an effective, even if opaque, semantic model to distinguish between legal (possible) behaviour and illegal (impossible) behaviour. Activities such as testing, fuzzing, or compiler optimisations fall within this broad range. 

A possible objection to our approach is that we generate training data sets from a known semantic model, whereas our stated initial problem referred to languages ``in the wild'' which have no such models. To respond, there is no substantial difference between generated plays from a known model and collecting interaction traces from instrumented real code, using some form of time-stamped profiling -- that is the consequence of the full abstraction result for the model. But this process is far more laborious then producing the traces from a known model. A known model has other advantages compared to using unknown code. Models of IA, both sequential and concurrent, have been studied algorithmically and are known to be complex, so the learning problem is non-trivial~\cite{DBLP:journals/tocl/Murawski05,DBLP:conf/tacas/GhicaM06}. For a controlled experiment in learnability ours is a suitable methodology, and the results indicate that applying the technique to real languages has potential.

\section{Idealised Algol (IA)}
For the sake of a focussed presentation we shall look at two variations on the programming language Idealised Algol (IA)~\cite{Reynolds1997}. IA is suitable for this experiment for several reasons. To begin with, it is a family of well-studied programming languages having at their core an elegant fusion of functional and imperative programming. We will concentrate in particular on two members of this family, Abramsky and McCusker's version of sequential IA~\cite{DBLP:journals/entcs/AbramskyM96} and Ghica and Murawski's version of concurrent IA~\cite{DBLP:journals/apal/GhicaM08}. Both these languages have mathematically precise (\textit{fully abstract}) game semantics which have an underlying common structure which makes it possible, but not trivial, to compare them. Finally, from a pragmatic point of view, the models themselves are elegant, can be presented concisely, and lend themselves well to computational experimentations. 

IA has basic data types, such is integers and booleans, with which three kinds of ground data types are constructed: commands (unit), local variables (references) and expressions. Function types are uniformly created out of these ground types. The terms of the language are those common in functional (abstraction and application, recursion, if expressions, arithmetic and logic) and imperative (local variables, assignment, dereferencing, sequencing, iteration). A peculiarity of IA, which sets it apart from most commonly encountered programming languages is the fact that it uses a \textit{call-by-name} mechanism for function application~\cite{DBLP:journals/tcs/Plotkin75}. For technical reasons, the IA we study here allows side-effects in expressions and admits a general variable constructor in which reading and writing to a variable can be arbitrarily overloaded. Concurrent IA, as described here, uses the same types plus a new type for \textit{binary semaphores}, along with new terms for parallel execution of commands and semaphore manipulation. 
Both languages have the type system of the simply-typed lambda calculus, with all language constants definable as (possibly higher-order) constants. 

\subsection{Game Semantics}

In game semantics the element of interaction between a term-in-context and the context is called a \textit{move}. Interactions characterising any particular execution are called \textit{plays}. All possible interactions with all possible contexts are called \textit{strategies.}

\newcommand{\qp}{\mathsf q}
\newcommand{\ap}{\mathsf a}
\newcommand{\op}{\mathsf o}
\newcommand{\pp}{\mathsf p}

Moves happen in \textit{arenas}, mathematical structures which define the basic causal structures relating such actions. An arena $A$ is a set $M$ equipped with a function $\lambda:M\rightarrow\{\op,\pp\}\times\{\qp,\ap\}$ assigning each move four possible \textit{polarities} which can be opponent/proponent and question/answer, and a relation ${\vdash}\subseteq M\times M$ called \textit{enabling}. Arenas interpret types. For example, the arenas for \textit{unit} and \textit{boolean} are:
\begin{align*}
\mathit{unit} &: M=\{q,a\},& & \lambda=\{(q,\op\qp),(a,\pp\ap)\},& & {\vdash}=\{(q,a)\} \\
\mathit{bool} &: M=\{q,t,f\},& & \lambda=\{(q,\op\qp),(t,\pp\ap),(f,\pp\ap)\},& & {\vdash}=\{(q,t),(q,f)\}.
\end{align*}
The significance of the \textit{question} $q$ is that a computation is initiated and of the \textit{answer} $a$ (or, respectively $t$ or $f$) is that a result is produced. The enabling relation establishes that the answer must be justified by the asking of the question. The interpretation of the opponent/proponent polarity is that ``proponent'' moves are initiated by the term whereas ``opponent'' moves by the context. As we can see, for computation at base types the computation is initiated via questions asked by the opponent, i.e. the context, and terminated via answers provided by the proponent, i.e. the term. The set of moves without an enabler are called \textit{initial} moves. In both examples above the set of initial moves is $I=\{q\}$.

From the basic arenas, \textit{composite} arenas may be created, for example for function type $A\Rightarrow B$ from arenas for $A$ and $B$:
\begin{align*}
	A &= \langle M_A, \lambda_A, {\vdash}_A \rangle\\
	B &= \langle M_B, \lambda_B, {\vdash}_B \rangle\\
	A\Rightarrow B &= \langle M_A\uplus M_B, \lambda_A^*\uplus\lambda_B, {\vdash}_A\uplus{\vdash}_B\uplus (I_B\times I_A) \rangle,
\end{align*}
where $\lambda^*$ is like $\lambda$ but with the $\op$ and $\pp$ polarities reversed. The significance of this polarity reversal is that in the case of arguments to a function the term/context polarity of the interaction becomes reversed. Enabling for function arenas relates not only moves in the two component arenas, but also each initial move in the argument $A$ to each initial move in the return type $B$, indicating that arguments may be invoked only after the function as a whole has started executing. 

For a term in context with type judgement $x_1:A_1,\ldots, x_k:A_k\vdash M:A$, with the arena in which it will be interpreted is that for $A_1\Rightarrow \cdots \Rightarrow A_k \Rightarrow A$. 

An interaction corresponding to an execution run of a term-in-context is called a \textit{play}, and it is a sequence of \textit{pointed moves} subject to correctness conditions which will be discussed later. A pointed move is an arena-move equipped with two \textit{names} (in the sense of~\cite{pitts2013nominal}), the first one representing its ``address'' in the sequence and the second one is ``the pointer'', i.e. the address of an enabling arena-move which occurs earlier in the sequence~\cite{DBLP:journals/entcs/GabbayG12}. 

For example, the typical play in the interpretation of sequential composition $\textit{seq}:\textit{unit}_3\rightarrow\textit{unit}_2\rightarrow\textit{unit}_1$ is
\[
q_1n_1\star\cdot
q_3n_2n_1\cdot
a_3n_3n_2\cdot
q_2n_4n_1\cdot
a_2n_5n_4\cdot
a_1n_6n_1.
\]
This sequence of actions is interpreted as: start computation ($q_1$), ask first argument ($q_3$, justified by initial question $n_1$), receive result ($a_3$, justified by preceding question), ask second argument ($q_2$, justified also by initial question $n_1$), receive result ($a_2$, justified by preceding question), indicate termination ($a_1$, justified by initial question $n_1$). Pointers are usually represented diagrammatically:
\begin{center}
	\includegraphics[]{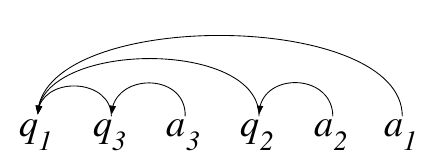}
\end{center}

Because the actual correctness conditions for plays are language-specific we will present them separately for sequential and concurrent IA, bearing in mind that everything up to this point is shared by the two. 

\subsection{Plays for sequential IA}

Given a justified sequence $s$ in an arena $A$ the notion of player and opponent \textit{view} are defined by induction as follows:

\begin{align*}
\mathit{pview}\ (\epsilon) &= \epsilon \\
\mathit{pview}\ (s\cdot mnn')&=\mathit{pview}(s)\cdot mnn' & \text{when }(\pi_1\circ\lambda) (m)=\pp\\
\mathit{pview} (s\cdot mn_1n_2 \cdot s'\cdot m'n_1'n_1)&=\mathit{pview}(s)\cdot mn_1n_2 \cdot m'n_1'n_1& \text{when }(\pi_1\circ\lambda) (m')=\op\\
\mathit{pview}\ (s\cdot mn\star)&=mn\star\\[2ex]
\mathit{oview}\ (\epsilon) &= \epsilon \\
\mathit{oview}\ (s\cdot mnn')&=\mathit{oview}(s)\cdot mnn' & \text{when }(\pi_1\circ\lambda) (m)=\op\\
\mathit{oview} (s\cdot mn_1n_2 \cdot s'\cdot m'n_1'n_1)&=\mathit{oview}(s)\cdot mn_1n_2 \cdot m'n_1'n_1& \text{when }(\pi_1\circ\lambda) (m')=\pp\\
\end{align*}

The view of a sequence is related to the stack discipline of computation in sequential IA, where certain actions, although present in the interaction traces, are temporarily ``hidden'' by other actions. 

The correctness conditions on sequences required for them to represent \textit{legal plays} are:
\begin{description}
	\item[alternation:] the proponent/opponent polarities of consecutive moves are different,
	\item[bracketing:] only the most recently unanswered question in a sequence can be answered,
	\item[visibility:] a proponent (opponent, respectively) move must have a justifier in the proponent (opponent, respectively) view of the preceding sequence 
\end{description}
Violating the alternation condition means that successive $\pp$-moves or $\op$-moves occur. The bracketing condition can be violated when questions are answered in the wrong order or multiple times:
\begin{center}
	\includegraphics[]{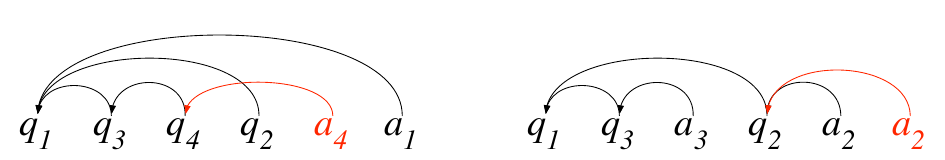}
\end{center}
Finally, the sequence below shows a violation of the visibility condition, for opponent:
\begin{center}
	\includegraphics[]{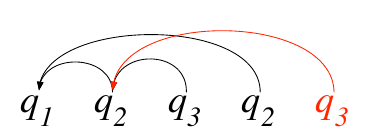}
\end{center}

\subsection{Plays for concurrent IA}

It is a general, and somewhat surprising, feature of game semantics that richer languages have simpler models. This is not as strange as it seems, because the more features a language has the more unrestricted its interaction with the context can be. In fact it is possible to think of ``omnipotent'' contexts in which the interactions are not constrained combinatorially~\cite{DBLP:journals/entcs/GhicaT12}. When sequential IA is enriched with parallelism, the alternation constraint disappears and bracketing and visibility are relaxed to the following, more general constraints:
\begin{description}
	\item[fork:] In any sequence $s\cdot qn_1n_1'\cdot s' \cdot mn_2n_1$ the question $q$ must be pending,
	\item[join:] In any sequence $s\cdot qn_1n_1'\cdot s' \cdot an_2n_1$ all questions justified by $q$ must be answered.
\end{description}
The idea is that a ``live thread'' signified by a pending question can start new threads (justify new questions) so long as it is not terminated (answered). Conversely, a thread can be terminated (the question can be answered) only after the threads it has started have also terminated. The simplest sequences that violate \textbf{fork} and \textbf{join}, respectively, are:
\begin{center}
	\includegraphics[]{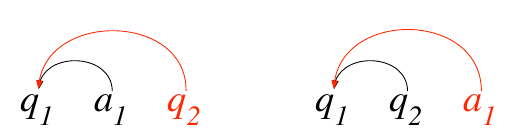}
\end{center}

For example, the typical play in the interpretation of parallel composition $\textit{par}:\textit{unit}_3\rightarrow\textit{unit}_2\rightarrow\textit{unit}_1$ is
\[
q_1n_1\star\cdot
q_3n_2n_1\cdot
q_2n_4n_1\cdot
a_3n_3n_2\cdot
a_2n_5n_4\cdot
a_1n_6n_1.
\]
This sequence of actions is interpreted as: start computation ($q_1$), ask first argument ($q_3$, justified by initial question $n_1$), immediately ask second argument ($q_2$, justified also by initial question $n_1$), receive the results in some order ($a_3$, justified by $n_1$, and $a_2$, justified by $n_4$), indicate termination ($a_1$, justified by initial question $n_1$). The play is represented diagrammatically as:
\begin{center}
	\includegraphics[]{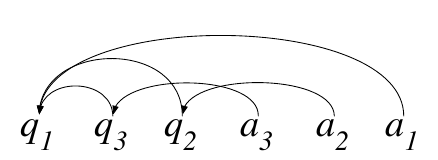}
\end{center}

\subsection{Strategies}

As we mentioned, plays characterise an interaction between term and context occurring in a particular run. In order to characterise the term we take the set of all such possible interactions, noting that they are also characterised by various \textit{closure conditions}. They all share prefix-closure as a common feature, typical to all trace-like models. In the case of sequential IA, the strategies are required to be \textit{deterministic} whereas in the case of concurrent IA they must be closed under certain permutations of moves in plays. It is strategies which give the fully abstract model of the language. 

We are not going to give the detailed definitions here because we shall focus on the learning of plays, rather than strategies. Learning strategies seems a more difficult proposition, which we will shall leave for future work. 

\subsection{Algorithmic considerations}\label{sec:algo}

Compared to the complexity of the syntax, the formal rules describing legality of behaviours in the language in terms of combinatorial properties of pointer sequences are remarkably succinct: just three rules for sequential IA (alternation, bracketing, and visibility) and two for concurrent IA (fork and join). A reasonable question to ask is whether these sets of sequences, taken as formal languages, are computationally complex or simple. 

It turns out that the answer depends on the order of the arena, where ground type is order 0 and an arena $A\Rightarrow B$ is the maximum between the order of $A$ plus one and the order of $B$. Plays in sequential IA defined in arenas of order up to 2 are regular languages, definable in terms of finite state automata~\cite{DBLP:journals/tcs/GhicaM03}, and for arenas of order up to 3 they are context-free languages, definable in terms of push-down automata~\cite{DBLP:conf/lics/Ong02}. Beyond this, strategies form undecidable languages~\cite{DBLP:journals/tcs/MurawskiW08}. In the case of concurrent IA, games in arenas of order 2 or more are undecidable~\cite{DBLP:journals/tcs/GhicaMO06}.

Of course, the results above refer to questions of language equivalence. Checking whether a (finite) sequence is a valid play in the models of sequential or concurrent IA is always decidable. But the results above suggest that the problem of learning such models is computationally challenging. 

\section{Learnability of IA models}

We will evaluate the learnability of sequential and concurrent IA using latent semantic analysis. First, a type signature is chosen, which determines an arena. Then a neural network is trained with random plays of the arena so that the level of \textit{perplexity} exhibited by the model is optimised. Using perplexity as a measure of accuracy is common in natural language processing. Given a probability model $Q$, one can evaluate it by how well it predicts a separate test sample $x_1,\ldots,x_N$ from a sample $P$. The perplexity of the model is defined by:
\begin{align*}
\Psi = 2^{-\frac{1}{N}\sum_{i=1}^N\text{log}_2Q(x_i)}.
\end{align*}
Better models have lower perplexity, they are less ``perplexed'' by the sample. In natural language processing, the perplexity of large corpora (1 million words) is of around 250 (per word). The exponent in the definition of perplexity is called the \textit{cross-entropy} and it indicates how many bits are required to represent the sequence in the word. For high quality natural language corpora, the cross-entropy is around 8 bits/word or 1.75 bits/letter~\cite{DBLP:journals/coling/BrownPPLM92}. The models are validated by computing the perplexity of a model against a different random sample of correct plays coming from the same language, over the same arena. A successful learning model will exhibit similar perplexities between the training set and the validation set. 

The accuracy of the learned model is then tested in two ways. The first test is to expose the model to a new sample, coming from the same programming language but perturbed using several single-character edits (insertions, deletions or substitutions) applied randomly to each sequence. This results in a set of plays at a small normalised Levenshtein distance from the correct plays used for training. Concretely, we use a distance of up to 0.1, e.g. 5 random edits applied to a sequence of length 50. In order for the test to be successful we expect to see a significant increase in perplexity between the test set as compared against the validation and training sets.

The second test is to expose the model to a sample of correct plays coming from the other language, i.e. testing the sequential model against concurrent plays and \textit{vice versa}. Noting that each sequential program is a particular (degenerate) form of a concurrent program, we expect the concurrently-trained neural network to exhibit similar levels of perplexity when exposed to the test data set and the validation data set, but we expect the sequentially-trained program to exhibit greater perplexity when exposed to the test data set --- since obviously there are concurrent plays which have no sequential counter-part. 

Game models are determined by the arena in which they happen. As discussed in Sec.~\ref{sec:algo}, the order of the arena has a significant impact on the algorithmic complexity of the model. We would expect games in low-order arenas to be faster to learn than games in higher-order arenas, but it is difficult to guess the effect the arena shape has over the accuracy of the model. As a consequence we examine both ``narrow'' and ``wide'' arenas. If we visualise and arena as a tree, the order is the height. The width of the arena corresponds to the number of arguments a function takes, and determines the number of distinct moves in its vocabulary of symbols. Below we show several arenas, depicted as trees:
\begin{center}
\includegraphics[]{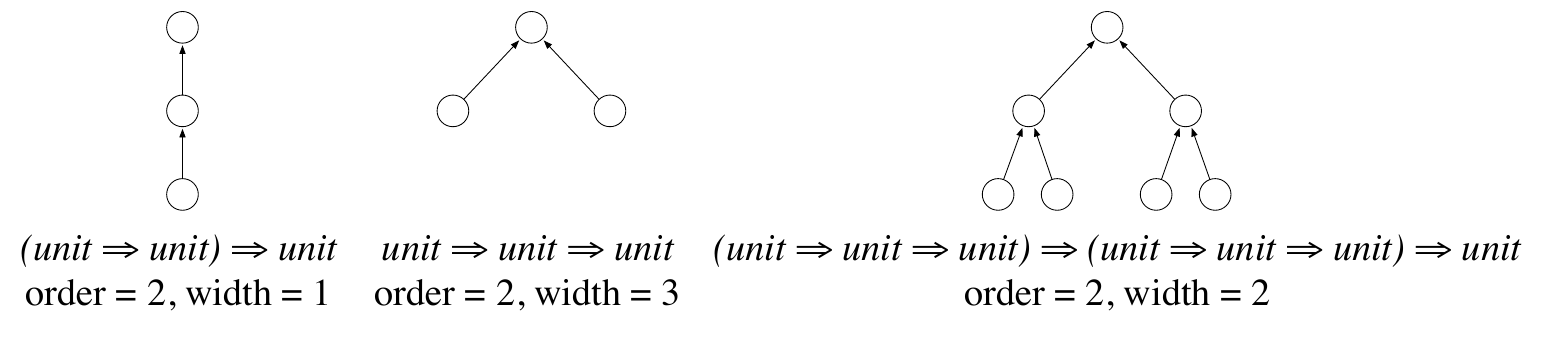}
\end{center}
The total number of moves is of the order $O(w^o)$ where $w$ is the width of the arena and $o$ the order of the arena. From the point of view of the language modelled, the width corresponds to the number of free variables in the term or the number of arguments a function might have, whereas the height is the order of the type of the term. We will conduct the experiment on arenas of orders 1 to 3. Going beyond 3 seems rather irrelevant as functions of order 4 or higher are rarely used in practice. We will conduct the experiment on arenas of width 1, i.e. functions taking one argument, in order to emphasise the complexity of the model as caused by higher-order features, and on arenas of width 5, i.e. functions with relatively large numbers of arguments. The two will be contrasted and compared. 

The corpora of plays we are creating for these arenas will be simplified further in two ways. 
The first simplification is that we replace all ground types with the \textit{unit} type. Indeed, in the legality rules for plays, sequential or concurrent, values play no role, and they can be safely abstracted by a generic notion of answer-move. This is important, because the presence of integers in plays would explode the vocabulary of moves beyond what is manageable. The second simplification is eliding the pointer information and focussing on sequences of moves only. This is a more significant and perhaps potentially controversial simplification. By eliminating pointers we make the model easier to learn, but we simultaneously make it less powerful since the pointer information is lost. The reason for removing the pointers is similar to that of removing values: they are usually represented by integers, and the presence of integers in traces increases too much the size of the vocabulary. Eliding pointer information is a common abstraction in games-based static analysis of programs~\cite{DBLP:conf/lics/GhicaB09}.

Pointers raise an additional problem in that the concrete representation matters~\cite{DBLP:journals/entcs/GabbayG12}, and it may clutter the learning process with extraneous information. For example, the sequence $qnn'\cdot an'n''$ (diagrammatically, \raisebox{-.25ex}{\includegraphics{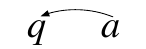}}) can be concretely represented, using integers for names, as $q01\cdot a12$, but also equivalently as $q10\cdot a03$. Alternative solutions could be considered, for example using a canonical representation of names, but as we will see the decision to simply abstract the pointers away is actually vindicated by the quality of the model. 

The length of random plays used in learning, validation and testing is at most 50. The size of the corpus of random sequences used for learning is 10,000 and 100,000, and the size of the corpora used for validation and testing are of 10,000 sequences. These parameters are arbitrary and not too important. Since the sequences are generated there are no limits on maximal sequence length or corpus size. Keeping in mind that the length of a play represents the number of function-argument interactions, a size of 50 seems generous. The number of plays in the corpora only impacts accuracy (which is already very good, as it will be seen) and the duration of the training process (which is reasonable, as it will be seen). For learning we use LSTMs, briefly described in Sec.~\ref{sec:lstm}. The details of the implementation and the (hyper)parameters of the model are discussed in the next section. 

\subsection{Detecting perturbed plays}

\begin{figure}
	\begin{subfigure}{0.5\textwidth}
		\includegraphics[width=1\textwidth]{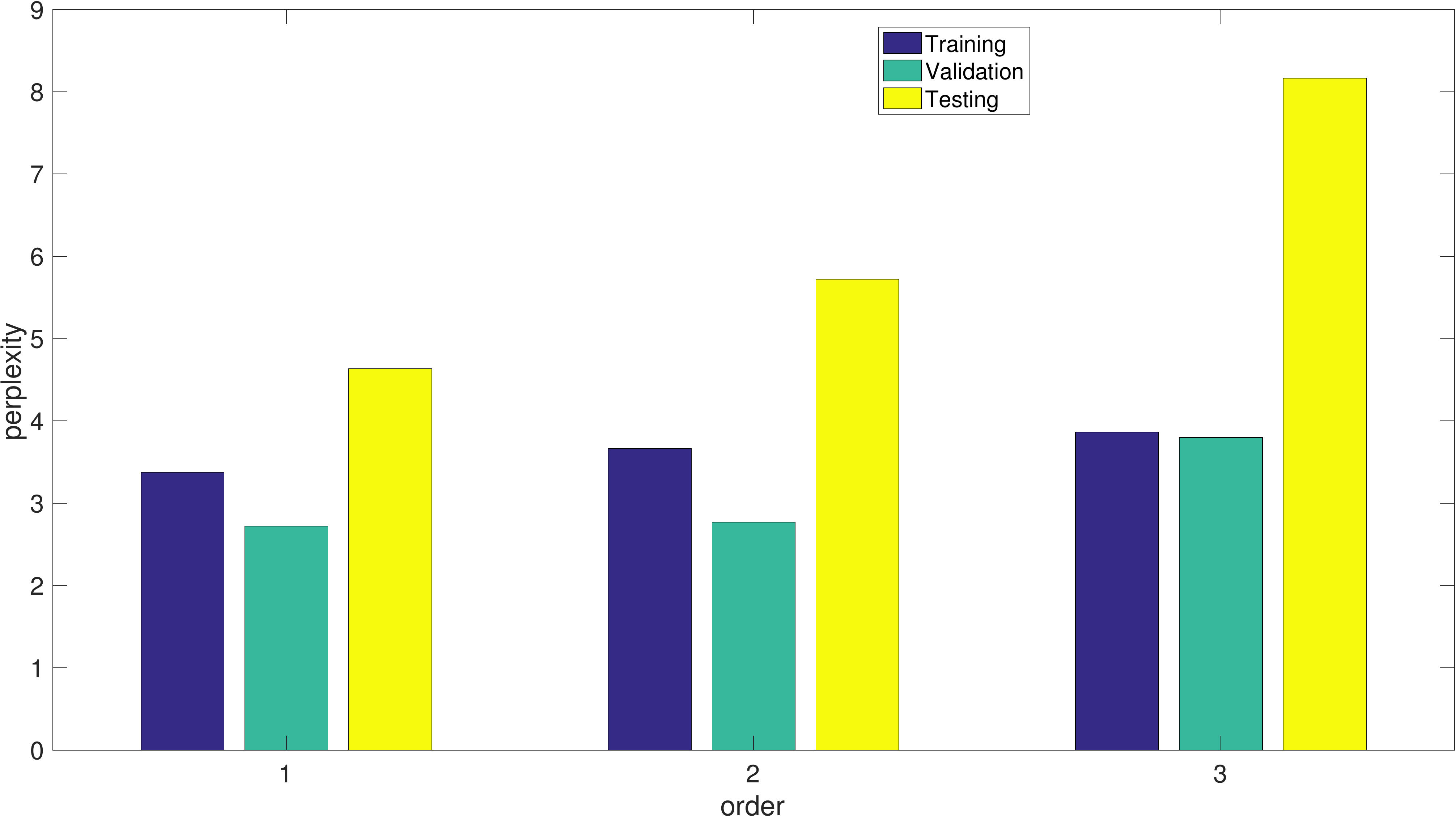}
		\caption{Width 1}
	\end{subfigure}  
	\begin{subfigure}{0.5\textwidth}
		\includegraphics[width=1\textwidth]{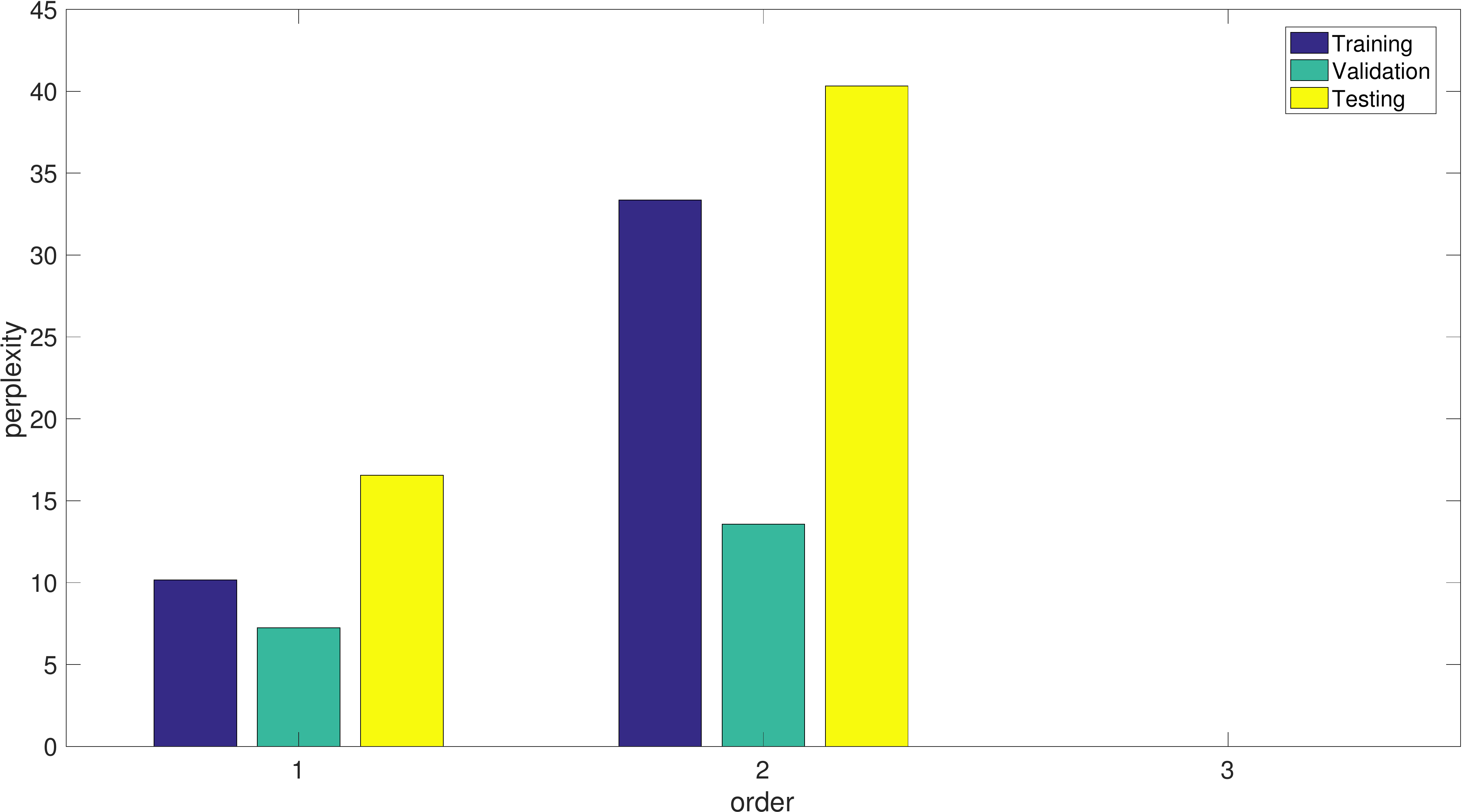}    
		\caption{Width 5}
	\end{subfigure} 
	\caption{Latent semantic analysis of perturbed sequential plays (10,000 plays)}
	\label{fig:learn10ks}
\end{figure}

\begin{figure}
	\begin{subfigure}{0.5\textwidth}
		\includegraphics[width=1\textwidth]{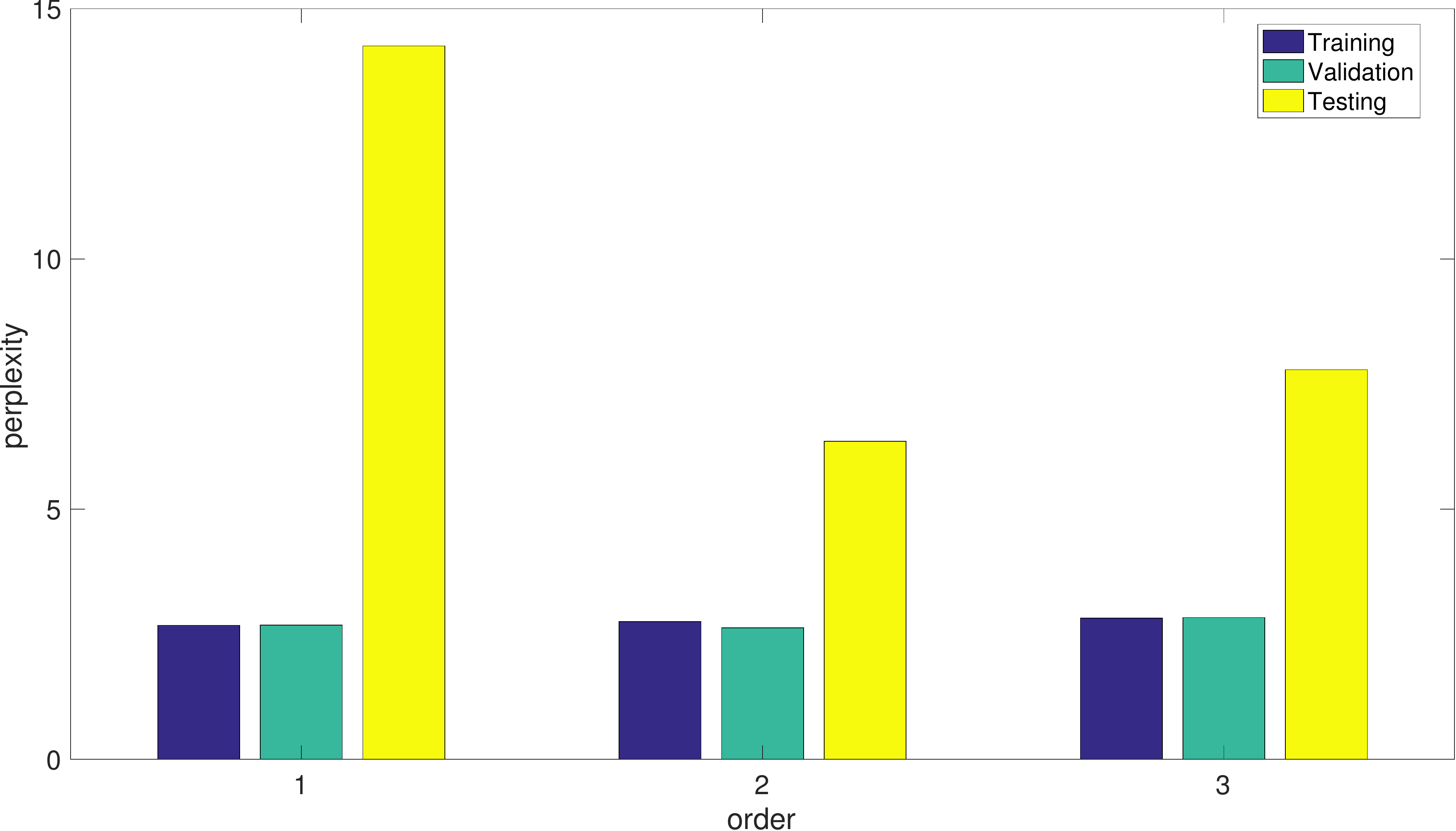}
		\caption{Width 1}
	\end{subfigure}  
	\begin{subfigure}{0.5\textwidth}
		\includegraphics[width=1\textwidth]{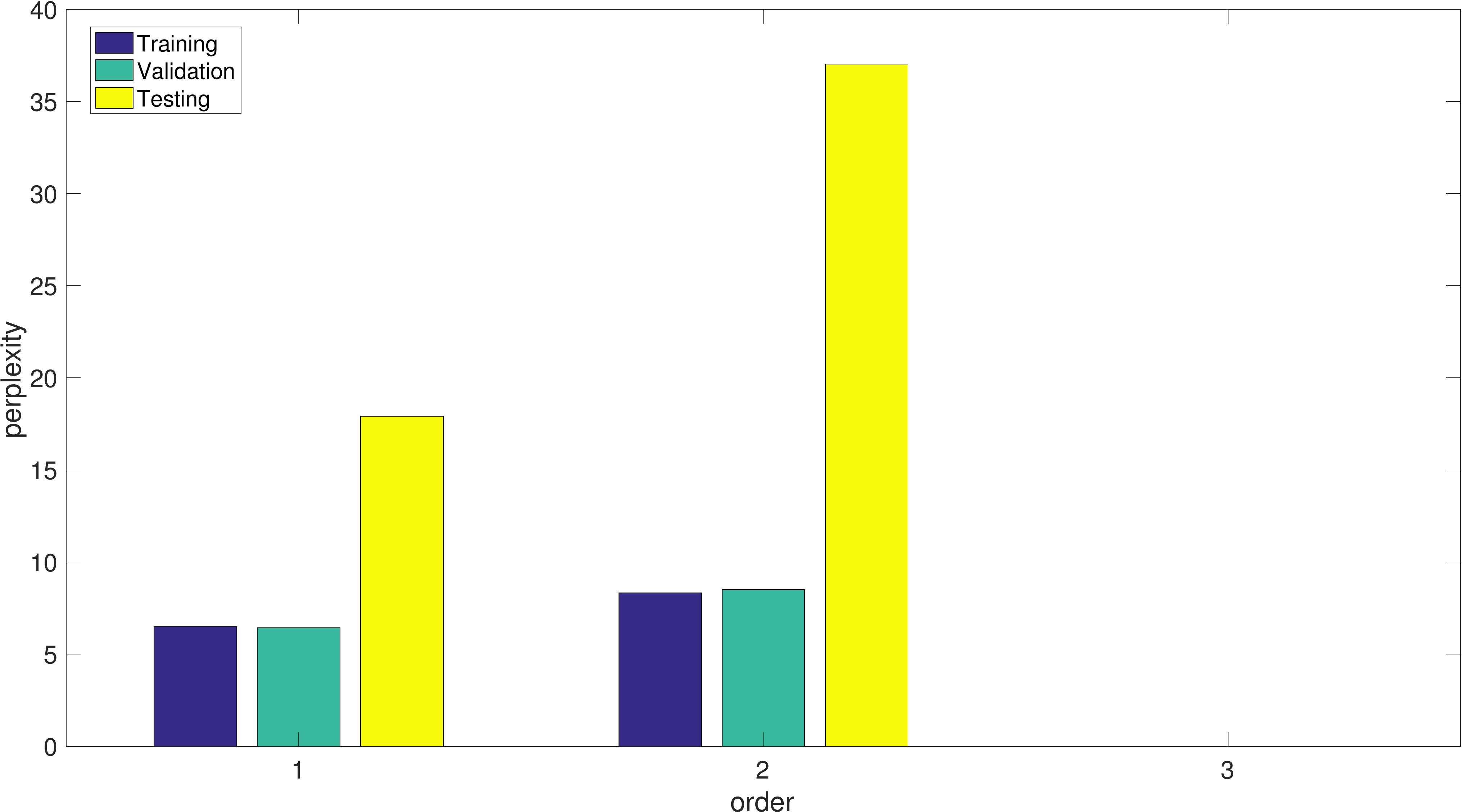}    
		\caption{Width 5}
	\end{subfigure} 
	\caption{Latent semantic analysis of perturbed sequential plays (100,000 plays)}
	\label{fig:learn100ks}
\end{figure}

\begin{figure}
	\begin{subfigure}{0.5\textwidth}
		\includegraphics[width=1\textwidth]{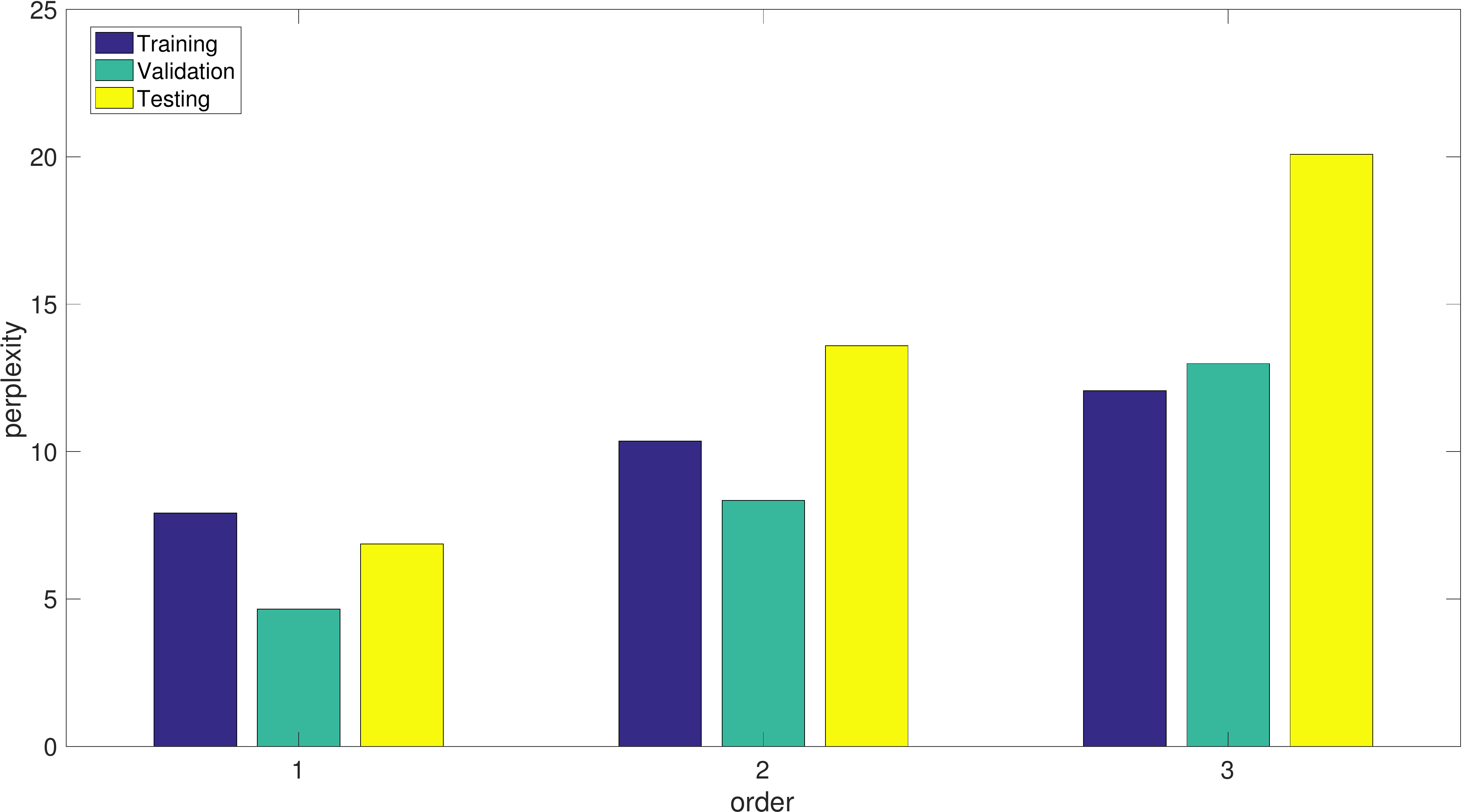}
		\caption{Width 1}
	\end{subfigure}  
	\begin{subfigure}{0.5\textwidth}
		\includegraphics[width=1\textwidth]{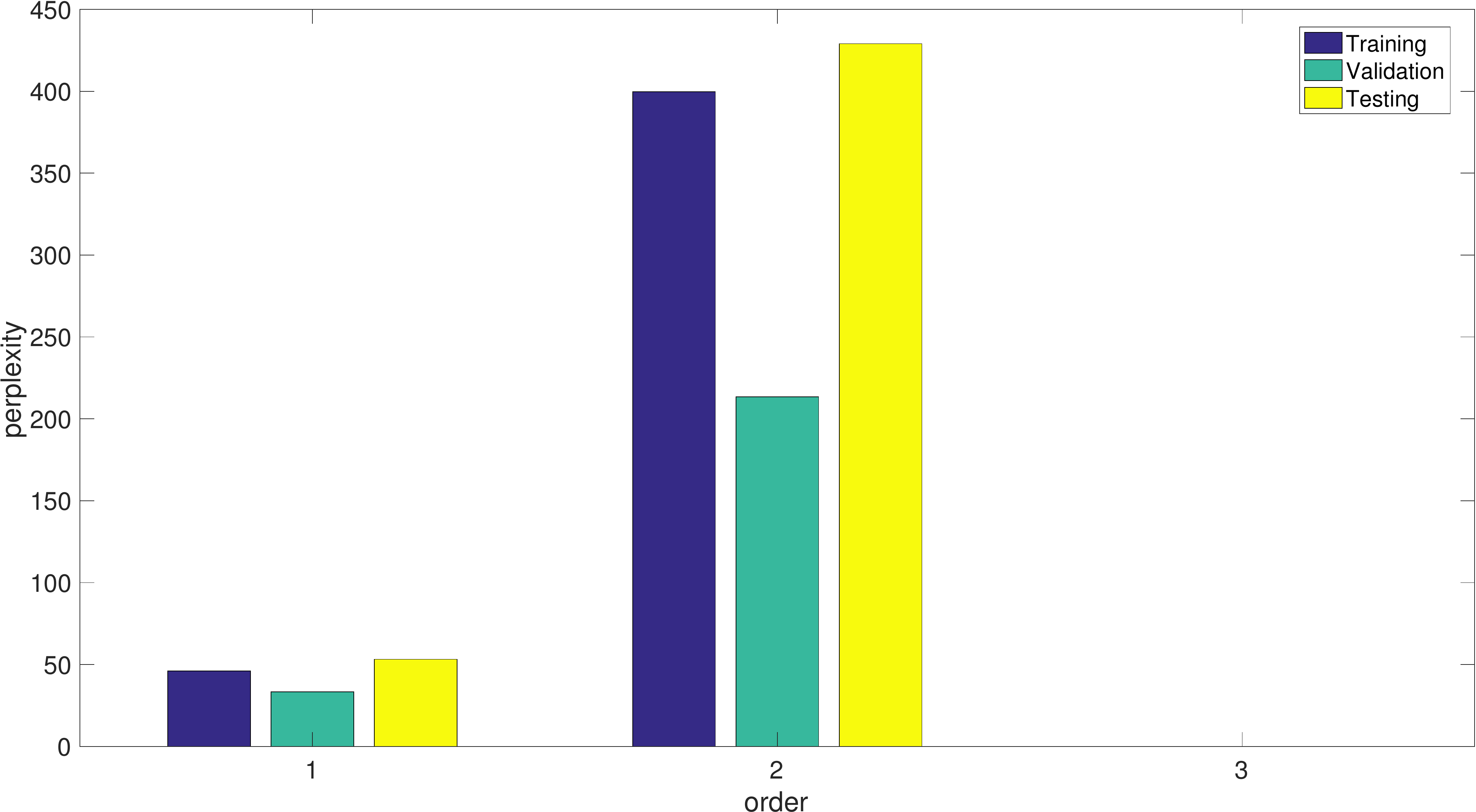}    
		\caption{Width 5}
	\end{subfigure} 
	\caption{Latent semantic analysis of perturbed concurrent plays (10,000 plays)}
	\label{fig:learn10kc}
\end{figure}

\begin{figure}
	\begin{subfigure}{0.5\textwidth}
		\includegraphics[width=1\textwidth]{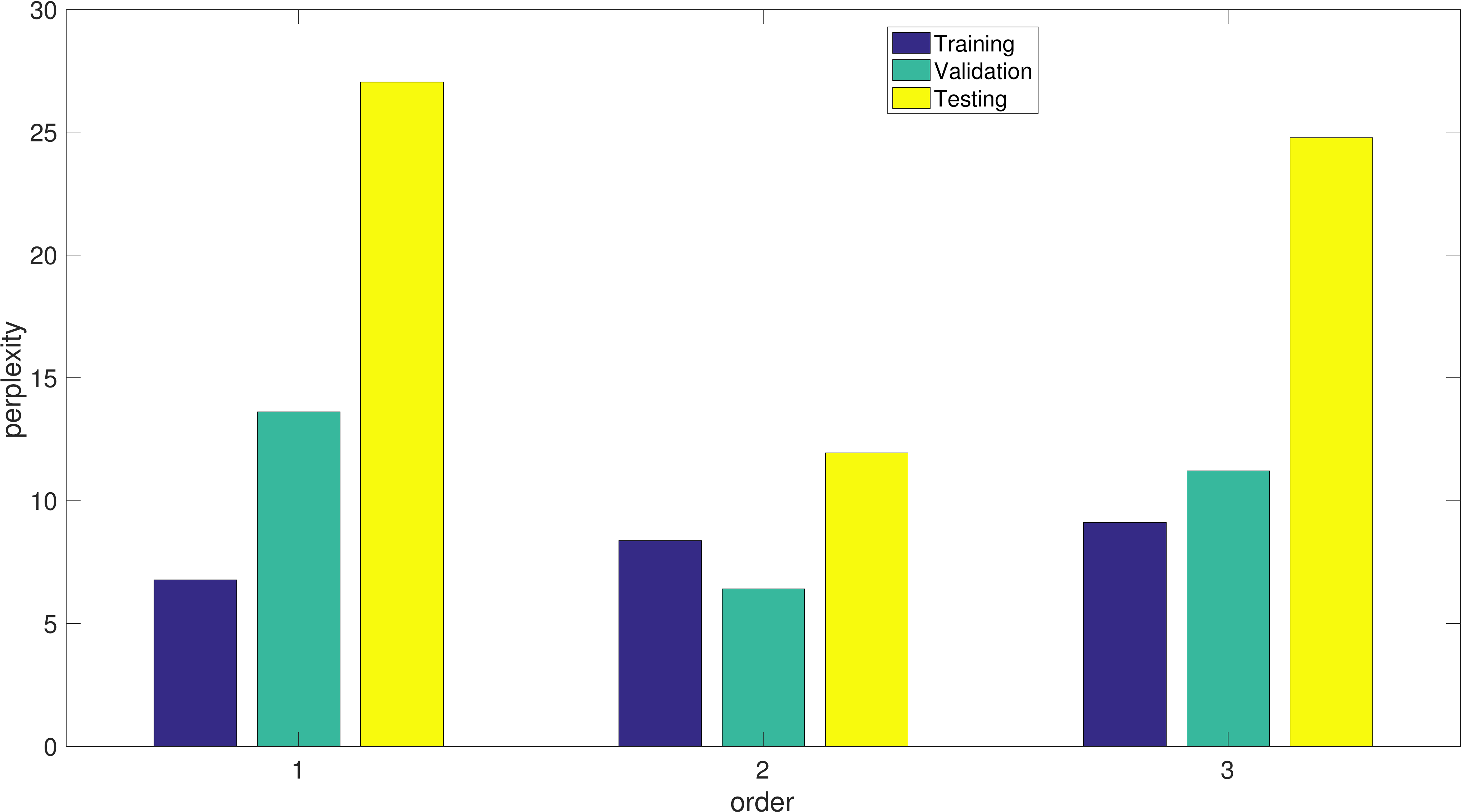}
		\caption{Width 1}
	\end{subfigure}  
	\begin{subfigure}{0.5\textwidth}
		\includegraphics[width=1\textwidth]{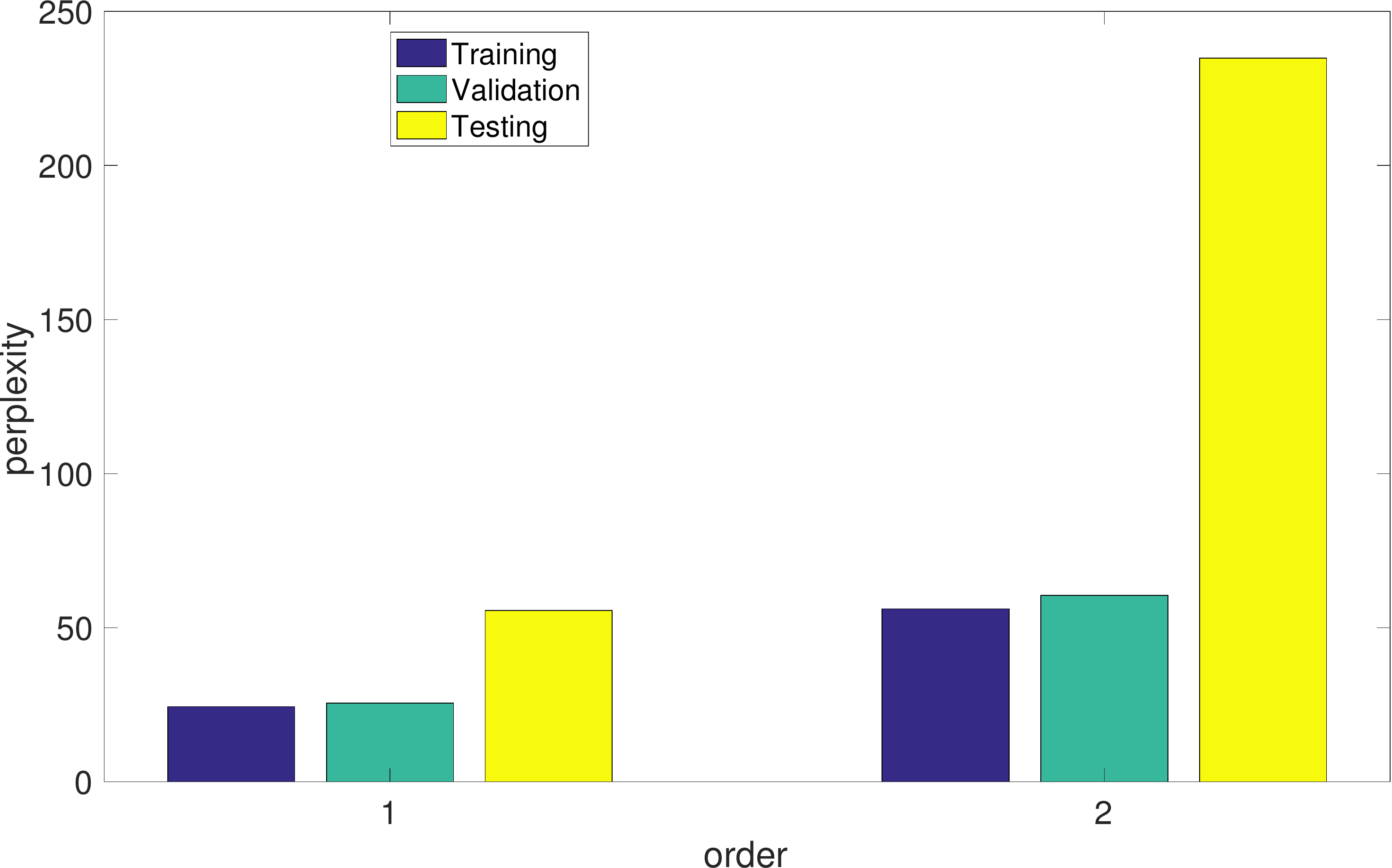}    
		\caption{Width 5}
	\end{subfigure} 
	\caption{Latent semantic analysis of perturbed concurrent plays (100,000 plays)}
	\label{fig:learn100kc}
\end{figure}

The first test involves applying the learned model to random sets of plays from the same language and the same arena at random normalised Levenshtein distance less than or equal to 0.1. The results are in Fig.~\ref{fig:learn10ks} for sequential models trained on 10K samples and in Fig.~\ref{fig:learn100ks} on models trained on 100K samples, and in Fig.~\ref{fig:learn10kc}-\ref{fig:learn100kc} for concurrent models. Each bar chart contains arenas of order 1-3. Where data is missing is because our hardware computational resources (memory) could not cope with the size of the model.

The first bar (navy) represents the perplexity of the model on the training data, the second (turquoise) the perplexity on validation data, and the third (yellow) the perplexity of the test data. In a good model the first two bars should be approximately equal and the third bar much higher. We note that in both the case of sequential and concurrent models the model is significantly more accurate when learned from 100k samples, rather than 10k samples. In absolute terms, the perplexity of the 100k samples models ranges from single digits to just over 50. However, the absolute perplexity is not relevant in latent semantic analysis, just the relative difference in perplexity between training, validation, and test data. In the worst cases (concurrent plays with width 1 order 5, and width 5, order 1) the difference is more than double, and in the best case more than 7 times, providing strong evidence for rejecting the set of Levenshtein-perturbed sequences. 

Experimental data strongly suggests that sequential games, which are more structured, are more learnable than concurrent games, which are looser. For sequential games the training and validation data perplexities are consistent, whereas the test data perplexity is much higher, at all arena orders and widths (100k samples). On the other hand, for concurrent games there are significant discrepancies between testing and validation perplexities, sometimes (width 1, order 1) by a factor of almost 2, suggesting that the model has not been learned as well as it could have been (maybe larger samples are required, maybe larger networks).

\subsection{Detecting different concurrency, sequentiality}

\begin{figure}
	\begin{subfigure}{0.5\textwidth}
		\includegraphics[width=1\textwidth]{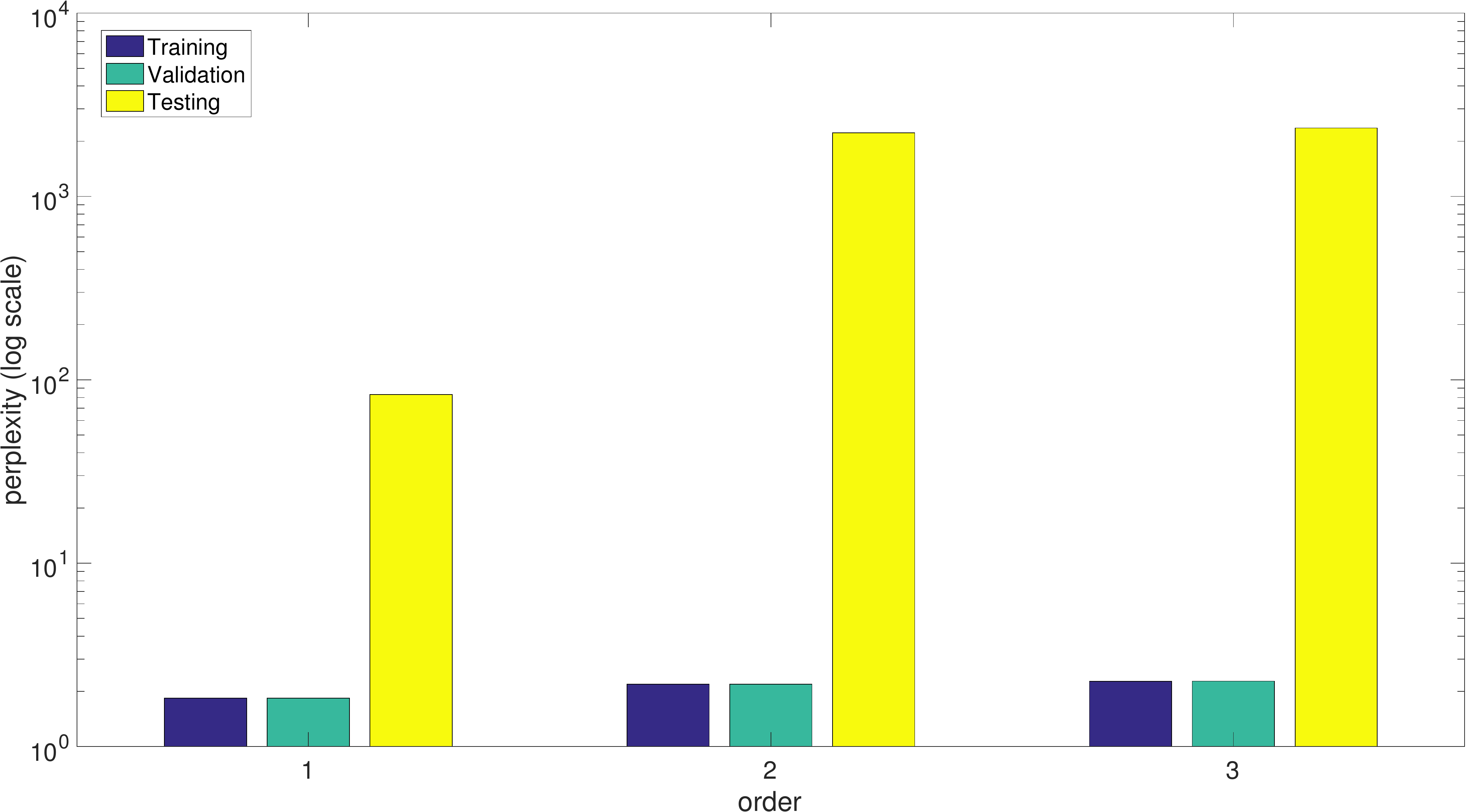}
		\caption{Width 1}
	\end{subfigure}  
	\begin{subfigure}{0.5\textwidth}
		\includegraphics[width=1\textwidth]{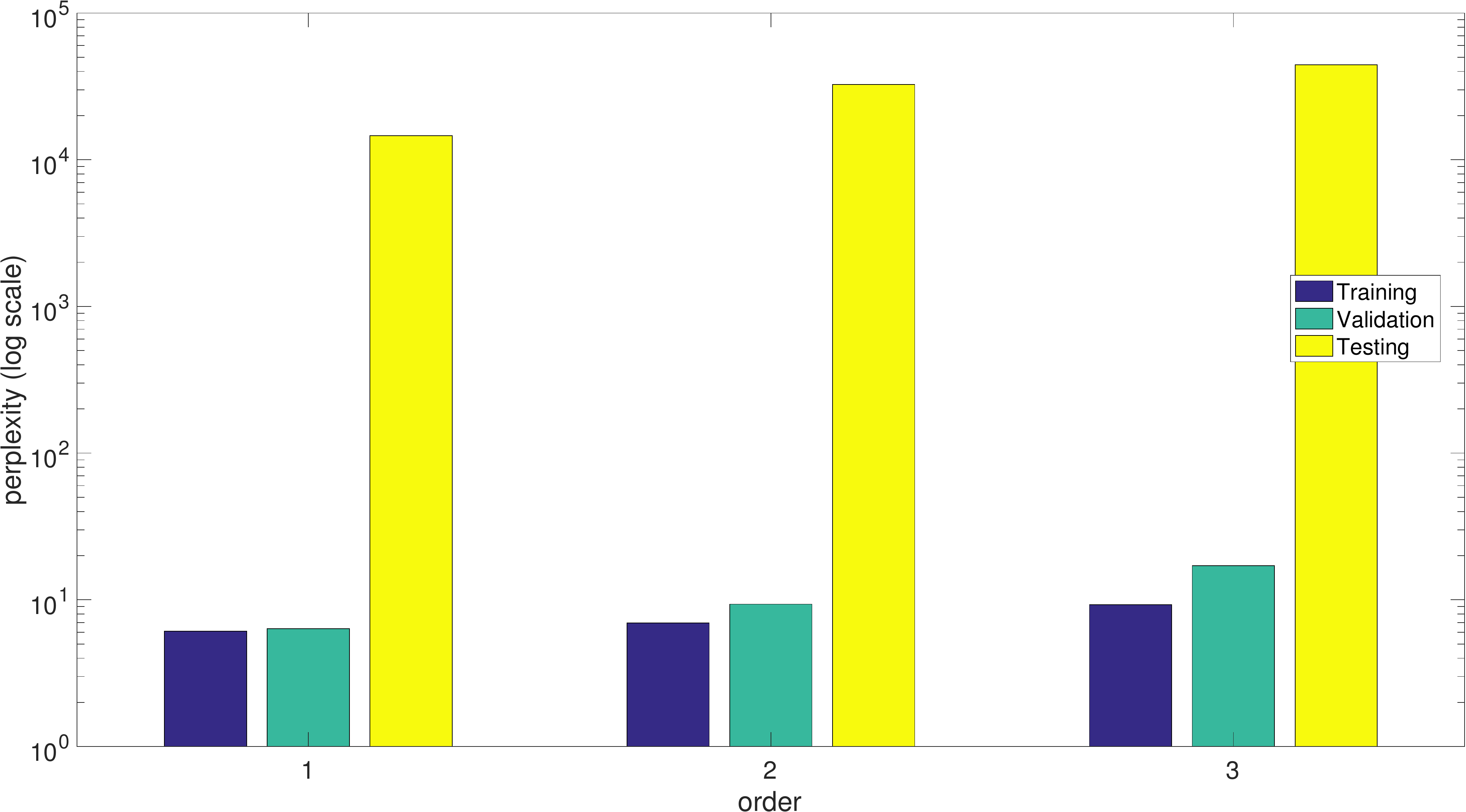}    
		\caption{Width 5}
	\end{subfigure} 
	\caption{Latent semantic analysis of concurrent plays in sequential models (10,000 plays)}
	\label{fig:learn10kst}
\end{figure}

\begin{figure}
	\begin{subfigure}{0.5\textwidth}
		\includegraphics[width=1\textwidth]{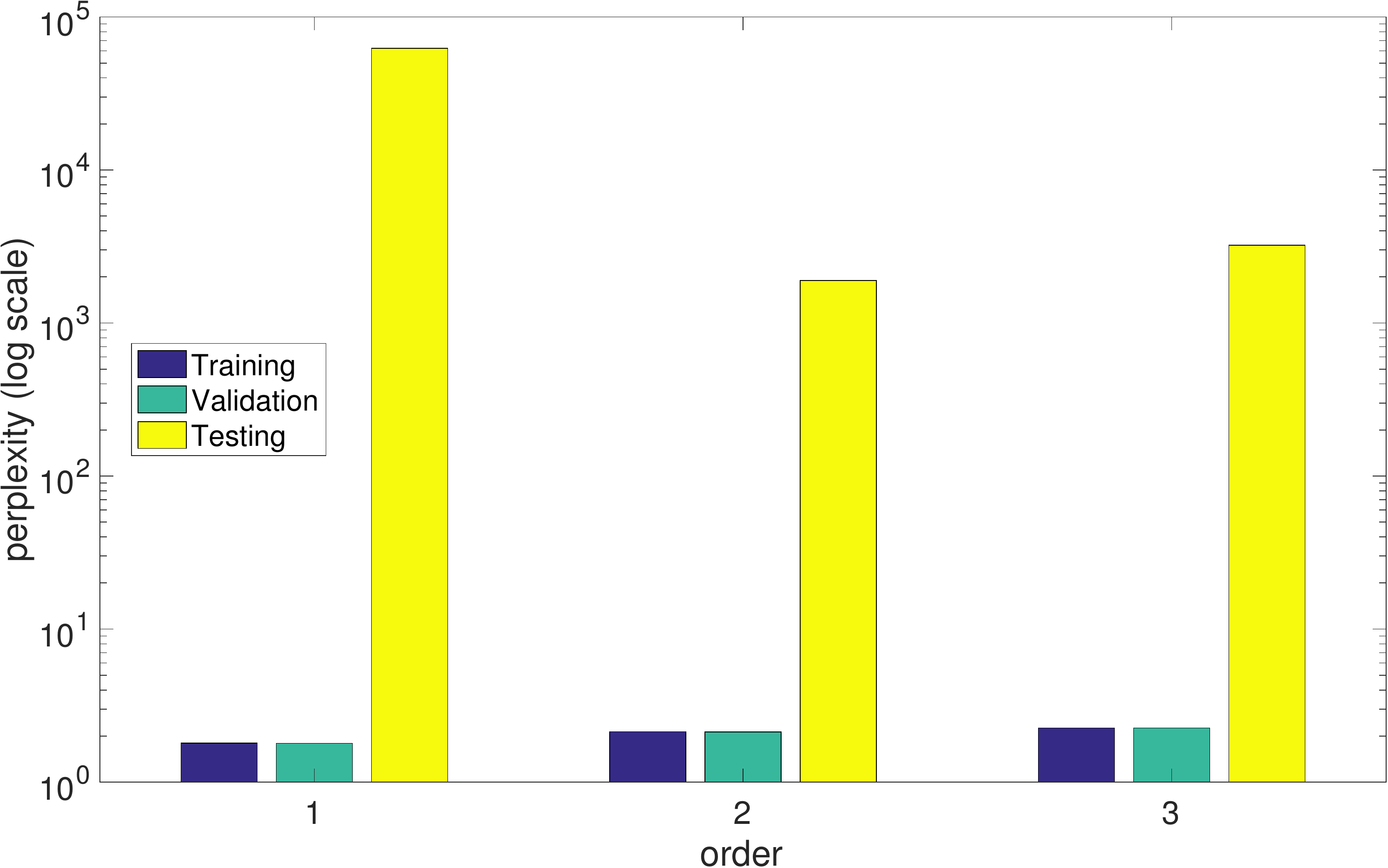}
		\caption{Width 1}
	\end{subfigure}  
	\begin{subfigure}{0.5\textwidth}
		\includegraphics[width=1\textwidth]{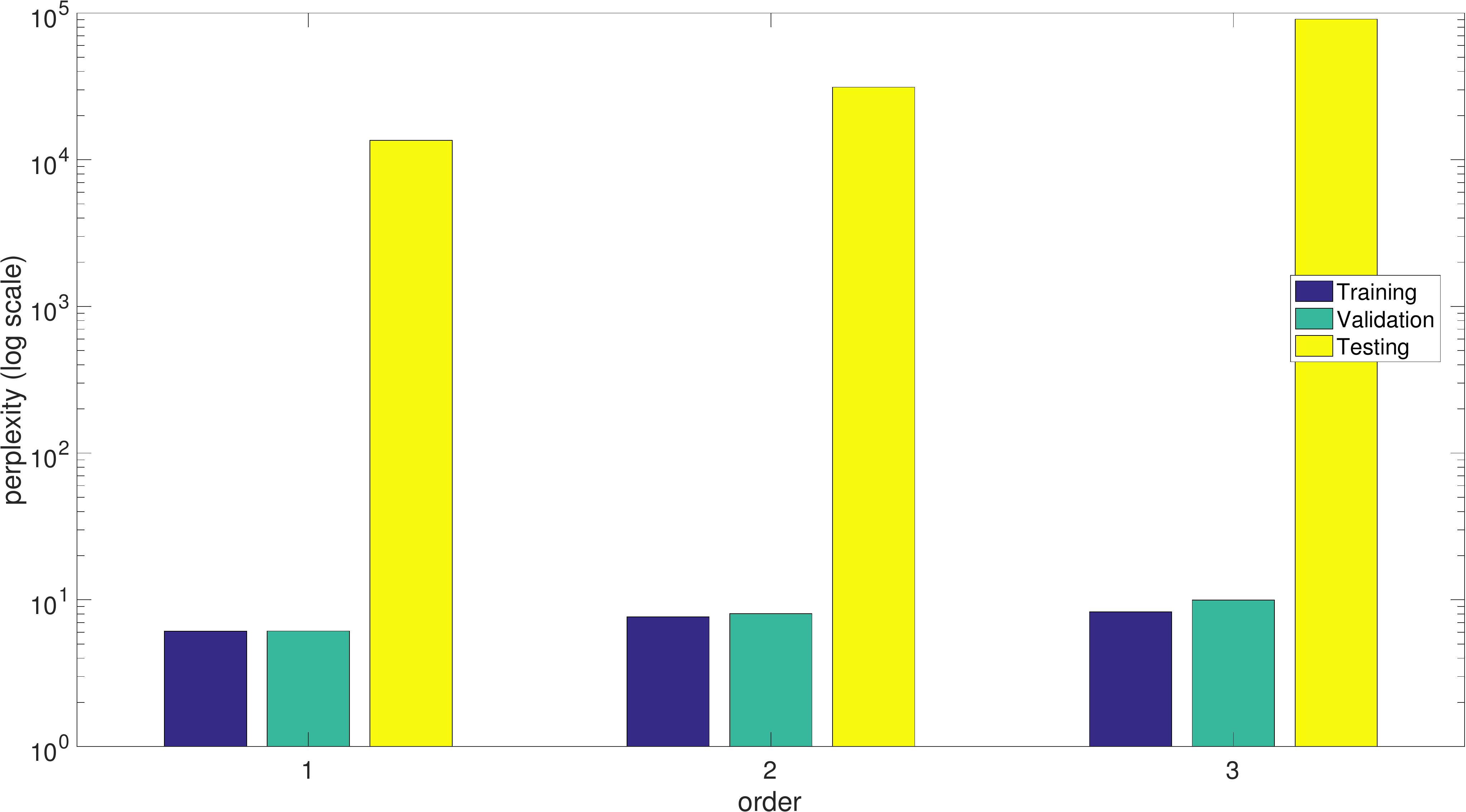}    
		\caption{Width 5}
	\end{subfigure} 
	\caption{Latent semantic analysis of concurrent plays in sequential models (100,000 plays)}
	\label{fig:learn100kst}
\end{figure}

The same models resulting from the learning process above have been used to perform latent semantic analysis of sequences provided by the other model. Fig.~\ref{fig:learn10kst} and~\ref{fig:learn100kst} show the perplexity (third bar, yellow) of testing a sequential model on concurrent plays over the same arena. In this case the evidence is overwhelming, from 2-5 orders of magnitude in perplexity increase. The extra training provided by using 100k samples is no longer relevant. 


\begin{figure}
	\begin{subfigure}{0.5\textwidth}
		\includegraphics[width=1\textwidth]{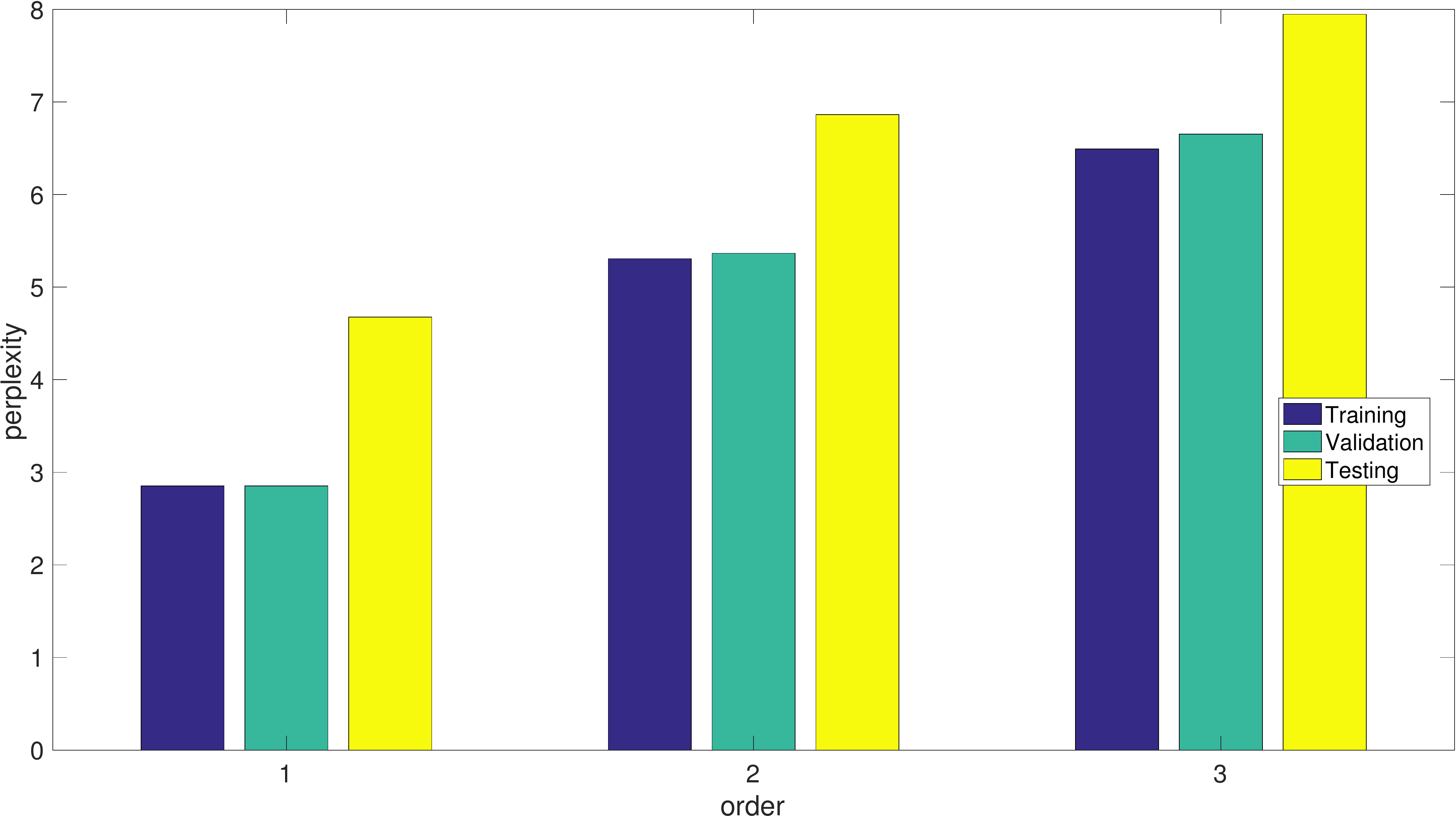}
		\caption{Width 1}
	\end{subfigure}  
	\begin{subfigure}{0.5\textwidth}
		\includegraphics[width=1\textwidth]{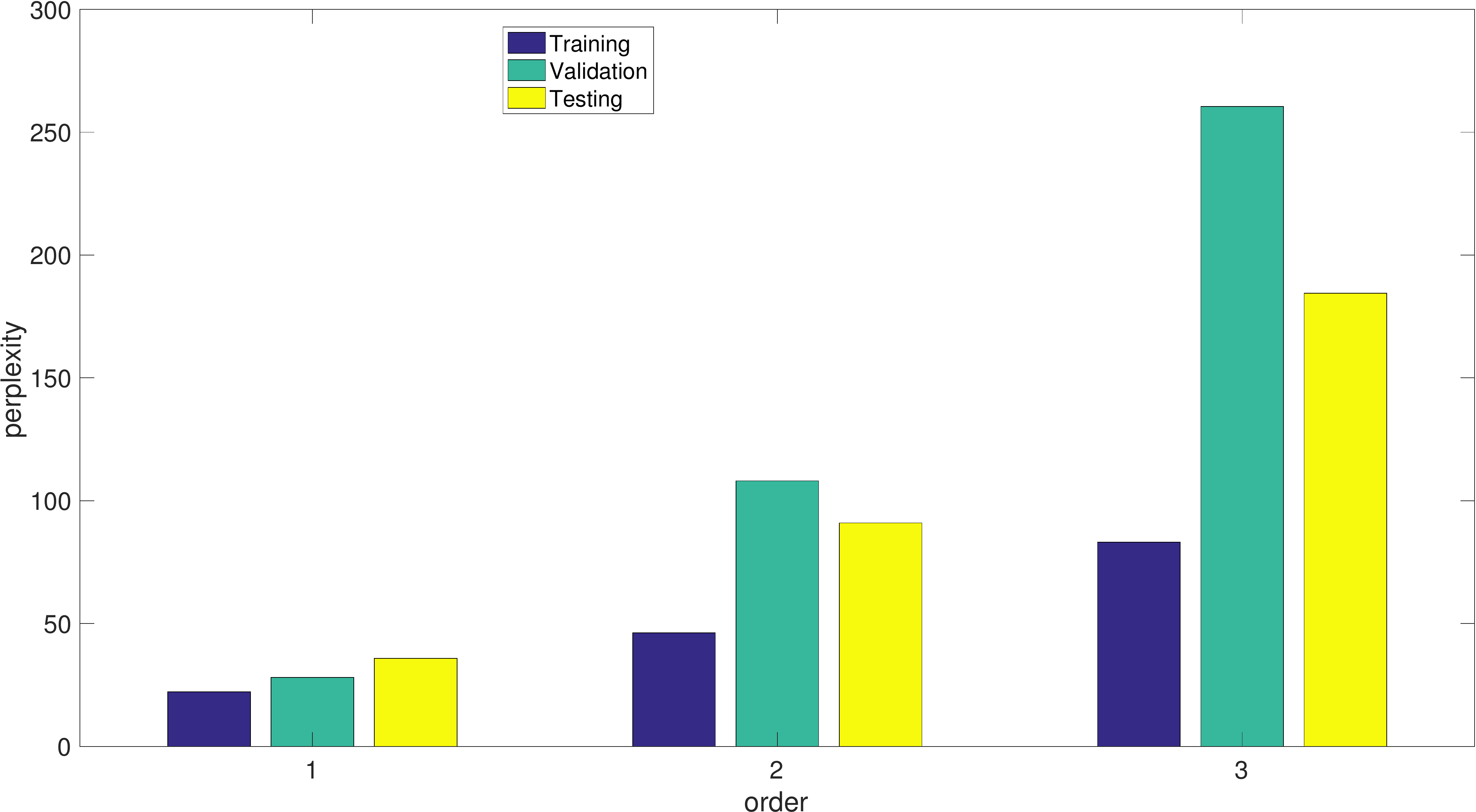}    
		\caption{Width 5}
	\end{subfigure} 
	\caption{Latent semantic analysis of sequential plays in concurrent models (10,000 plays)}
	\label{fig:learn10kct}
\end{figure}

\begin{figure}
	\begin{subfigure}{0.5\textwidth}
		\includegraphics[width=1\textwidth]{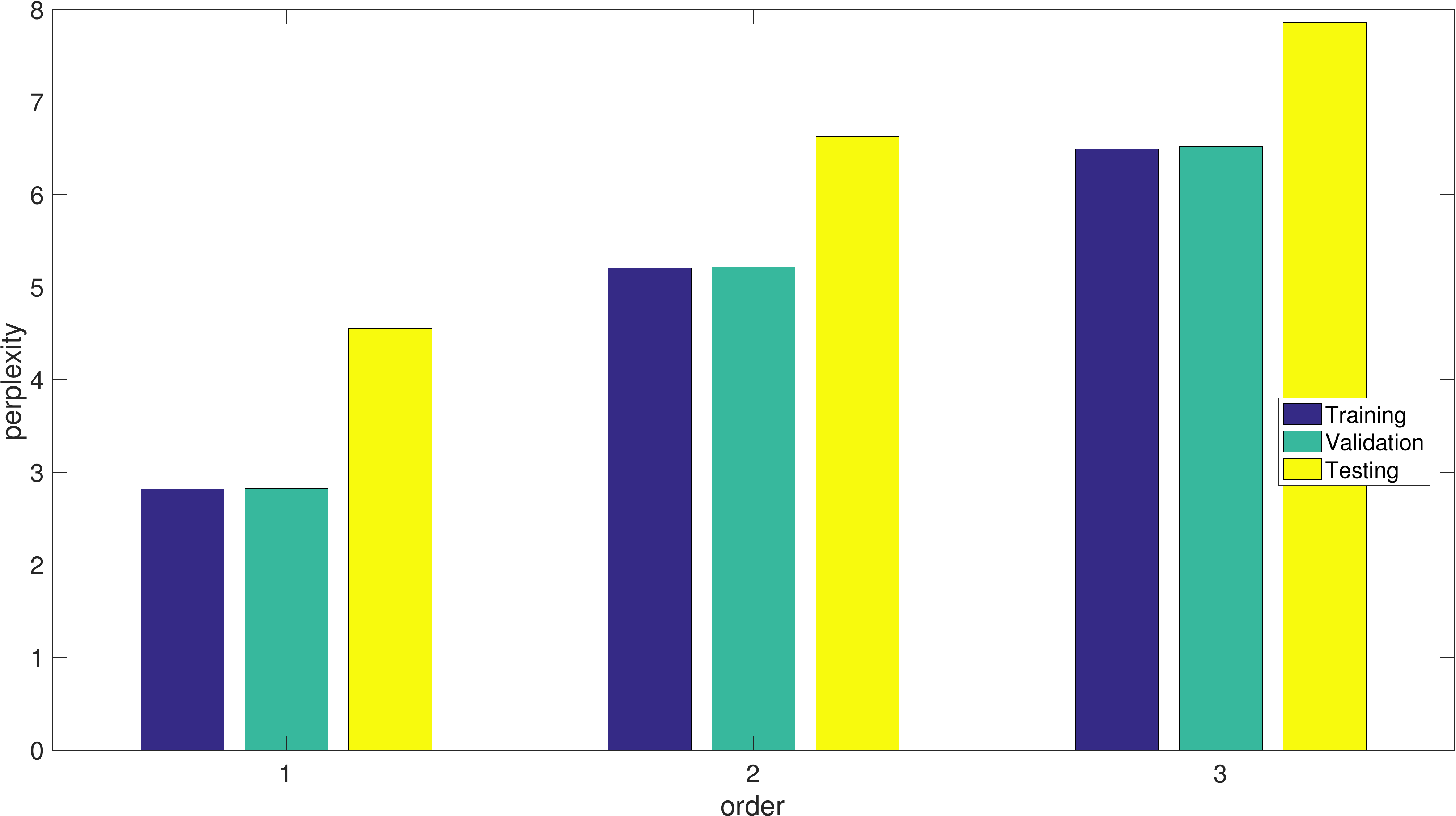}
		\caption{Width 1}
	\end{subfigure}  
	\begin{subfigure}{0.5\textwidth}
		\includegraphics[width=1\textwidth]{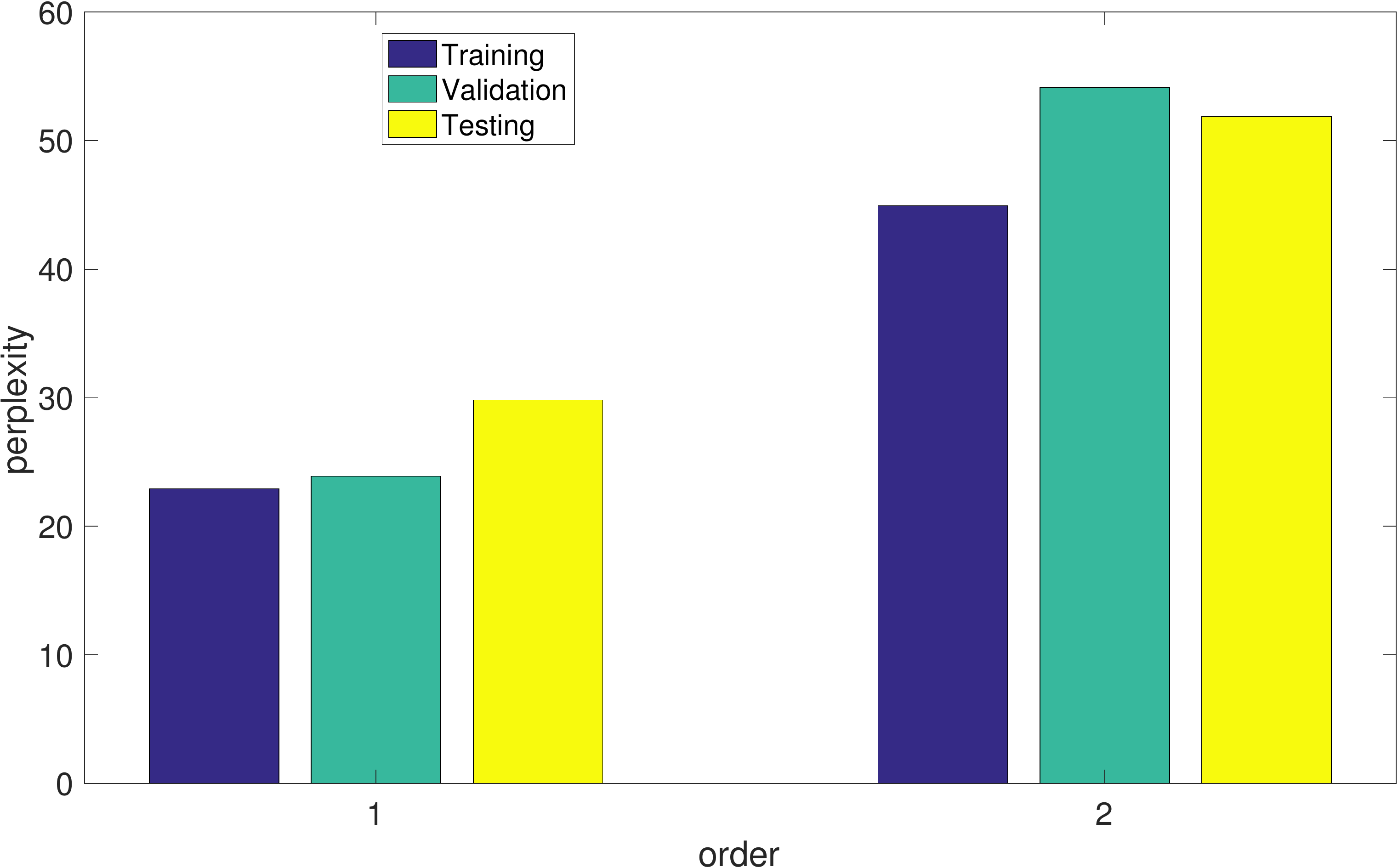}    
		\caption{Width 5}
	\end{subfigure} 
	\caption{Latent semantic analysis of sequential plays in concurrent models (100,000 plays)}
	\label{fig:learn100kct}
\end{figure}

Finally, using the concurrent model to test sequential plays (Fig.~\ref{fig:learn10kct}-\ref{fig:learn100kct}) shows, as expected the concurrent model cannot identify sequential plays, since all sequential plays can be found in the concurrent model. The perplexity of the test set is not as low as that in the validation set, and that is not entirely surprising because of the emphasis of the learning process. Sequential sequences are relatively rare among all concurrent sequences, so it is reasonable for their perplexity to be higher. 


\subsection{Implementation notes}

We are using the standard implementation of LSTM distributed with \textsc{TensorFlow}\footnote{\url
{https://github.com/tensorflow/models}}. The model uses an LSTM cell which processes moves sequentially, computing probabilities for possible values of the next move in the sequence. The memory state is initially zeroes, updated after each word. Ideally, in a recurrent neural net (RNN), the output depends on arbitrarily distant inputs. However, this makes the training process computationally intractable, so it is common in practice to ``unfold'' the net a fixed number of steps; in the concrete case of our model this value is 20. The inputs are represented using a dense embedding. This is considered undesirable for text but it is demanded here by the large size of the symbol set~\cite{DBLP:conf/acl/BaroniDK14}. The loss function for the model is the sample perplexity, discussed earlier. To increase the expressive power of the model, two LSTMs are layered, each containing 200 nodes. This is considered a small LSTM model. 

\begin{figure}
\centering
\includegraphics[scale=.3]{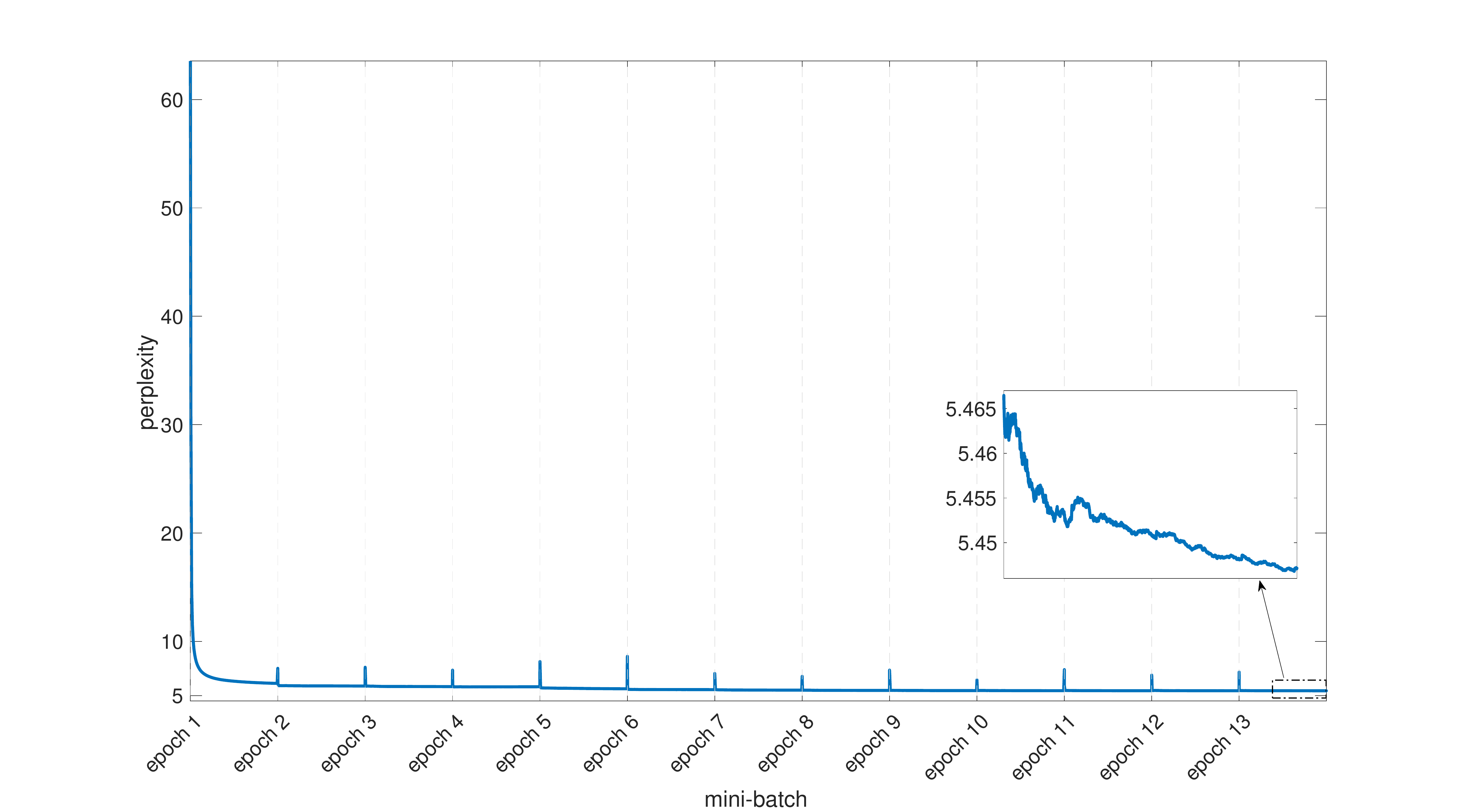}
\caption{Training convergence of the model on the perturbed sequential plays of order 2 and width 5. The perplexity is shown after processing each mini-batch (size 20) for all training epochs. The learning rate is fixed for the first 4 epochs at 1, and then reduced by a factor of 2 at each subsequent epoch reaching 0.002 in epoch 13.}
\label{fig:epoch}
\end{figure}

The training cycle consists of several (13) epochs, although in almost all cases except the largest arenas, the model converges after only 1-2 epochs of training. Further training leads to little or no improvement in the model, as seen in Fig.~\ref{fig:epoch}, which is a typical example. The experiments were carried out on a mid-range CUDA device, GeForce GTX 960. The training cycle for each model was around one hour. 

\section{Conclusion, related and further work}

\subsection{Recurrent neural nets}\label{sec:lstm}

A \textit{perceptron} is a simple computational element from a vector of real numbers to real numbers, which behaves like a weighted sum of the input composed with a step function. A perceptron is \textit{trained} by adjusting the weights and the threshold values so that it fits a given set of examples. A \textit{feed-forward neural network} (FFN) is essentially a directed acyclic graph in which each node is a perceptron. The most common graph topology for a FNN consists of several \textit{layers} of perceptrons so that each output from any given layer is connected to all inputs of the subsequent layer. A FFN is trained using \textit{back-propagation}, which is a family of gradient-descent algorithms for adjusting the weights and thresholds of the perceptrons to match a given training data set. 

Traditional FFNs have been successfully applied to many machine learning problems, however when it comes to the task of sequence-learning, the architecture of an FFN suffers from two main limitations: it cannot readily handle inputs of arbitrary length and it does not explicitly model time \cite{lipton2015critical}. Further more, FFN models that implement some form of a sliding context window to implicitly capture the time dependency between the inputs cannot sufficiently model the time since the range of the captured dependency is limited by the size of the window \cite{bengio2003neural,DeMulder2015}.

Recurrent neural networks (RNN), unlike FNNs, allow for the presence of cycles in their underlying topology. This creates memory-like effects in the network which allow dynamic temporal behaviour. The way in which the RNN is topologically structured is connected to both its expressiveness and the training algorithms. As a result of these compromises, RNN architectures can be very diverse. 

Unlike FNNs, recurrent neural networks (RNNs) can readily handle inputs of arbitrary length and can model the temporal patterns present in sequential data. The main difference in the architecture of RNNs and FFNs is the addition of recurrent edges in RNNs, which span adjacent time steps, introducing a notion of ``memory". RNNs have immense expressive power, a finite-sized RNN with non-linear activation function is a universal approximator \cite{SCHAFER2007} and can simulate a universal Turing machine \cite{siegelmann1991turing}. Capturing time dependencies is not only limited to RNNs, for example traditional statistical models such as Hidden Markov models (HMMs) \cite{Stratonovich1960} capture the these dependencies through modelling an observed sequence as an output of a sequence of unobserved states, however, they also typically fail to capture long-range dependencies. This drawback stems from two reasons, namely the relative small size of the state space, and the short distant dependencies among hidden states, since the transition probabilities between states typically depend only on the immediate previous states. Increasing the state space size increases the time and space required for the training algorithm quadratically \cite{Viterbi1967}, while increasing the dependency distance between hidden states (corresponding to larger context window) increases both of the space and time exponentially \cite{lipton2015critical}. This therefore renders HMMs  computationally infeasible for modelling long-range dependencies. 

On the other hand, the expressive power of an RNN grows exponentially with the number of its hidden nodes while the growth of the training complexity is maintained to a polynomial (at most quadratic growth) \cite{lipton2015critical}. The number of distinct states that a hidden layer of an RNN can represented grows exponentially with the number of nodes in the layer. RNNs are applied to a wide range of machine learning problems that involve sequence learning such as language modelling \cite{Mikolov2012,DeMulder2015,jozefowicz2016exploring}, machine translation \cite{sutskever2014sequence}, speech recognition \cite{Graves2013}, and image caption generation \cite{xu2015show}. In most of sequence learning tasks, an RNN or a variant of it is usually the state-of-the-art method. RNNs have been applied not only to learning natural languages, but also artificially generated languages of algorithmic patterns, and proved themselves to be more effective than other methods~\cite{joulin2015inferring}.

The addition of the recurrent edges to the architecture of RNNs gives them great expressive powers, however, they also introduced the ``vanishing and exploding gradient'' problem which occurs while training the network when the errors are back-propagated across many time steps \cite{Bengio1994}. The Long Short-Term Memory (LSTM) is a crucial variant of RNNs that was introduced by \cite{DBLP:journals/neco/HochreiterS97} specifically to address this problem. Unlike conventional nets in which the weights have the role of an implicit and quite rudimentary memory, LSTMs have explicit memory cells in their architecture, used to store gradient information for training. The  architecture of the LSTM is quite sophisticated and a detailed presentation is beyond the scope here, but accessible tutorials are available~\cite{olah17}.

\subsection{Machine learning for programming languages}

Using machine learning for programming language semantics is largely new and unexplored terrain, even though heuristic search techniques such as genetic algorithms have been applied to software engineering problems~\cite{DBLP:journals/csur/HarmanMZ12}. This is a well researched area which is related but complementary to our interest. Primarily, search-based software engineering (SBSE) is a collection of syntactic techniques, which rely on manipulation of code, usually as a syntax tree, to extract information about the code, to manipulate the code, or to detect patterns in the code (common bugs, anti-patterns, etc.). There is a significant area of overlap between the aims and techniques of SBSE and other heuristic-heavy programming-language analyses and manipulation such as refactoring, slicing, test-generation, or verification. By contrast, semantic models are independent of syntax. In fact the kind of analysis we have proposed here ignores syntax and relies directly on program behaviour instead. Indeed, latent semantic analysis of code when the source code is available is trivial: one can merely scan it for occurrences of terms associated with concurrency, such as parallel execution or semaphores. The problem becomes more interesting when the source code is not available: given a piece of compiled code, e.g. a module or a library, can we determine whether it originates in one language or another just by examining the way it interacts with its calling context? Our analysis shows that at least sometimes the answer is positive. 

There are some obvious limitations to our approach. First of all we looked for distinctions in plays rather in strategies, just because learning strategies (potentially infinite sets of plays) seems significantly harder than learning the plays themselves. But there are semantic differences between languages which are only reflected at the level of the strategy. For example PCF~\cite{DBLP:journals/iandc/HylandO00}, sequential IA~\cite{DBLP:journals/entcs/AbramskyM96} and non-deterministic IA~\cite{DBLP:conf/lics/HarmerM99} have the same notion of legality on plays but differ at the level of strategies. PCF requires \textit{innocent strategies}, sequential IA \textit{deterministic strategies}, and non-deterministic IA \textit{non-deterministic strategies}. Moreover, the formulation of these distinctions requires both pointers and answer-values, information which we abstract away from our modelling. Using our set-up these distinctions are lost. Capturing such subtle distinctions would require a different approach. 

However, by and large the results of our experiment are very encouraging. The quality of the models is high, as evidenced by their robust discriminatory powers, and the required computational resources modest. These results make us optimistic about using this methodology on practical programming languages as encountered ``in the wild''. The process of creating a corpora of training traces in the absence of a model is of course different. We need a large code base, a compiler, and a way to instrument the interface between a part of the code taken to be as ``the term'' and the rest of the program taken to be ``the context''. Much like a profiler, the instrumentation should record how, in any execution, the term interacts with the context via its free variables (function or method calls and returns). 

What is interesting is that a model of code obtained from real code ``in the wild'' will learn not only what is ``legal'' behaviour but also what is ``idiomatic'' behaviour -- patterns of behaviour which are specific to the code-base used for learning. Depending on the quality of the model this can have  some possibly interesting applications. Note that this same phenomenon appears in the case of machine-learning natural language from corpora~\cite{Caliskan183}, except that in the case of programming languages the idiomatic aspects are more likely to be seen as embodiments of de-facto practices rather than problematic biases.

For example, the model can be used for novelty detection~\cite{DBLP:journals/sigpro/MarkouS03, DBLP:journals/sigpro/MarkouS03a} in order to augment code inspections: instead of merely studying the code syntactically, the behaviour of code-in-context can be analysed for conformance with the existing body of code. Syntax-independent novelty analysis can have other well-known applications, for example to security. Unexpected or unusual patterns of interactions are, for example, typical for attempts to compromise the integrity of a system. 

A recogniser running in reverse is a generator, and generating valid traces -- especially idiomatic valid traces -- is a possibly interesting way of automating the testing of functional interfaces. Generating random data for automating testing is a well-understood process~\cite{DBLP:conf/icfp/ClaessenH00}. However, generating random functional behaviour is a much more complicated proposition, and syntactic approaches do not seem equally promising. 

Semantic-directed techniques, in particular using models that are both compositional and operational such as trace semantics or game semantics, have been advocated for a long time but did not make as deep inroads as expected in the practice of programming. A pragmatic disadvantage is that semantic models can be mathematically demanding, but this is not even the main problem. The main difficulty is that on the balance they are both difficult to construct and brittle, in the sense that small changes to the language can require a total rethinking of its semantic model. Moreover, most languages are not syntactically (and semantically) self-contained because they interact with  other languages via mechanisms such as foreign function interfaces. Machine learning, if effective on real languages, solves both these problems. It hides the mathematical complexity of the model behind the automated learning, and it can derive models out of existing code-bases, capturing not only a sense of what is legal but also what is idiomatic (for engineering, but also cultural reasons) in a particular language. In the end, as it tends to be the case with machine learning, the resulting model may be opaque and uninformative but it may end up being  effective enough for practical purposes.
\\[1.5ex]
\noindent
\textbf{Acknowledgements.} This paper is motivated by a challenge from Martin Abadi. Preliminary experiments were conducted by Victor Patentasu and were presented at the \textit{Off the Beaten Track} workshop of POPL 2017. Khulood Alyahya has been supported by EPSRC grant EP/N017846/1. Dan R. Ghica has been supported by EPSRC grant EP/P004490/1.

\bibliographystyle{eptcs}

\end{document}